\documentclass[oldversion]{aa} % double column
\usepackage{graphicx}
\usepackage{longtable}
\usepackage{txfonts}
\usepackage{rotating}
\usepackage{lscape}
\newcommand{\gsim}{\;\lower.6ex\hbox{$\sim$}\kern-7.75pt\raise.65ex\hbox{$>$}\;}
\newcommand{\lsim}{\;\lower.6ex\hbox{$\sim$}\kern-7.75pt\raise.65ex\hbox{$<$}\;}
\newcommand{\teff}{$T_{\rm eff}$}

\begin{document}
\title{The Na-O anticorrelation in horizontal branch stars. V. NGC~6723
\thanks{Based on observations collected at 
ESO telescopes under programme 087.D-0230}
\fnmsep
\thanks{Tables 2, 3, 5, 6, and 7 are only available in electronic form}}

\author{
R.G. Gratton\inst{1},
S. Lucatello\inst{1},
A. Sollima\inst{2},
E. Carretta\inst{2},
A. Bragaglia\inst{2},
Y. Momany\inst{1,3},
V. D'Orazi\inst{4,5},
M. Salaris\inst{6},
S. Cassisi\inst{7,8}
\and
P.B. Stetson\inst{9}}

\authorrunning{R.G. Gratton}
\titlerunning{Na-O in HB stars of NGC~6723}

\offprints{R.G. Gratton, raffaele.gratton@oapd.inaf.it}

\institute{
INAF-Osservatorio Astronomico di Padova, Vicolo dell'Osservatorio 5, I-35122
 Padova, Italy
\and
INAF-Osservatorio Astronomico di Bologna, Via Ranzani 1, I-40127, Bologna, Italy
\and
European Southern Observatory, Alonso de Cordova 3107, Vitacura, Santiago, Chile 
\and
Department of Physics \& Astronomy, Macquarie University, Balaclava Rd., North Ryde, Sydney, NSW 2109, Australia
\and
Monash Centre for Astrophysics, School of Mathematical Sciences, Building 28, Monash University, VIC 3800, Australia
\and
Astrophysics Research Institute,
           Liverpool John Moores University,
           IC2, Liverpool Science Park,
           146 Brownlow Hill,
           Liverpool L3 5RF, UK
\and
INAF-Osservatorio Astronomico di Teramo, Via Collurania, Teramo, Italy
\and
Instituto de Astrofisica de Canarias, La Laguna, Tenerife, Spain
\and
Dominion Astrophysical Observatory, Herzberg Institute of Astrophysics, National Research Council, 
5071 West Saanich Road, Victoria, British Columbia V9E 2E7, Canada }

\date{}
\abstract{We used FLAMES+GIRAFFE (Medusa mode) at the VLT to obtain moderately high resolution 
spectra for 30 red horizontal branch (RHB) stars, 4 RR Lyrae variables, and 17 blue horizontal 
branch (BHB) stars in the low-concentration, moderately metal-rich globular cluster NGC~6723 
([Fe/H]=$-1.22\pm 0.08$\ from our present sample). The spectra were optimized to derive O and 
Na abundances. In addition, we obtained abundances for other elements, including N, Fe, Mg, Ca, 
Ni, and Ba. We used these data to discuss the evidence of a connection between the distribution 
of stars along the horizontal branch (HB) and the multiple populations that are typically 
present in globular clusters. We found that all RHB and most (13 out of 17) BHB stars are 
O-rich, Na-poor, and N-poor; these stars probably belong to the first stellar generation in 
this cluster. Only the four warmest observed stars are (moderately) O-poor, Na-rich, and N-rich, 
and they probably belong to the second generation. While our sample is not fully representative 
of the whole HB population in NGC~6723, our data suggest that in this cluster only HB stars 
warmer than $\sim 9000$~K, that is one fourth of the total, belong to the second generation,
if at all. Since in many other clusters this fraction is about two thirds, we conclude that the 
fraction of first/second generation in globular clusters may be strongly variable. In addition, 
the wide range in colour of chemically homogeneous first-generation HB stars requires a 
considerable spread in mass loss ($>0.10$~M$_\odot$). The reason for this spread is yet to be 
understood. Finally, we found a high Ba abundance, with a statistically significant radial 
abundance gradient.}

\maketitle

\section{Introduction}

The horizontal branch (HB) of globular clusters (GCs) is the locus in their 
colour-magnitude diagram (CMD) that contains the stars that are burning He in their cores. 
The initial location of the stars on the HB (the zero-age horizontal branch, ZAHB) is 
expected to depend primarily on their mass and chemical composition; while 
still burning helium in their cores, the stars experience limited changes in surface 
temperature and luminosities; they become much redder and brighter only in very late 
stages of the HB phase when very little He is left in their cores. Until recently, 
GCs were considered to be simple stellar populations, that is, all stars were thought
to have the same age and chemical 
composition. Following that scheme, we expect that HB stars have a colour 
distribution due only to their evolution off the ZAHB, unless they loose different 
amounts of mass while still on the red giant branch (RGB). However, it was immediately
clear from the first quantitative comparison between models and data several 
decades ago that, in each GC, stars distribute over a range in colour much wider 
than expected from this simple evolutionary consideration, with considerable variation 
in this range for different clusters (see, e.g., Rood 1973). In addition, it is also known that median colours 
of the HB stars in different clusters - originally assumed to be coeval - are not 
simply a function of metallicity, as originally expected based on evolutionary models 
(see Faulkner 1966). These two facts constitute the "second parameter" problem (Sandage 
\& Wildey 1967; van den Bergh 1967). Many different explanations have been suggested for 
this phenomenon (for reviews, see Fusi Pecci \& Bellazzini 1997 and Catelan 2009). In 
the past few years, it has become clear that most of the complex phenomenology can be 
explained by a combination of different factors: cluster-to-cluster age differences, 
metallicity-dependent mass loss with small but not negligible star-to-star 
variations, and star-to-star variations in the He abundances related to the multiple 
population phenomenon (Gratton et al. 2010; see also Dotter 2013 and Milone et al. 2014). 
In addition, other factors (binarity, variation of the total CNO content, and perhaps 
rotation; see e.g. Rood 1973) 
should also be considered, at least for some specific clusters. Moreover, the relative 
role of the star-to-star variations in mass loss and in chemical composition is yet 
to be determined. 

This complex problem can be attacked following a variety of approaches, including
statistical analyses of large samples of GCs or studies of the properties of the variable 
stars, for instance. Basic information can be obtained by determining the chemical composition of 
stars in different locations along the HB. In fact, we expect that evolved He-rich stars 
(belonging to the so-called second generation in GCs, SG; Gratton et al. 2012a) have 
masses lower than He-normal ones (members of the first generation, FG), and they should
occupy a bluer location on the HB (see Ventura et al. 2001; %Bedin et al. 2004; Norris 2004;  
D'Antona et al. 2002; Piotto et al. 2005; Cassisi et al. 2013, following the earlier 
suggestions by, e.g., Rood 1973 and Norris et al. 1981). When the He abundances of 
the individual HB stars are known, synthetic HB populations can be compared with 
observed distributions of stars along the HB, allowing determination of the 
residual scatter in mass that can be attributed to other factors.

While conceptually simple, this method has several limitations. He abundances are 
difficult to derive directly because He lines are very weak in cool stars 
($T_{\rm eff}<9000$~K) and because stars warmer than the so-called Grundahl jump 
(at about $T_{\rm eff}\sim 11,500$~K; Grundahl et al. 1999) do not have an outer 
convective envelope. In such warm stars, sedimentation and radiative levitation 
strongly affect the composition of the atmosphere, making it not feasible to
deduce the original stellar composition. On the other hand, we may use elements 
other than He as useful diagnostics of the multiple population phenomenon for 
stars cooler than the Grundahl jump. Very useful information can be obtained using 
Na and O lines: while the relation between the abundances of these elements and 
that of He is probably not linear (Ventura \& D'Antona 2008; Decressin et al. 2007), 
large He abundances in SG stars are always accompanied by a large overabundance of 
Na and deficiency of O with respect to the composition of He-normal FG stars of the 
same cluster. Stars on the blue end of the HB of a cluster are then expected to be 
richer in Na and N, and poorer in O, than stars at the red end.

First attempts to use this approach for a better understanding of the HB morphology have
been the studies by Villanova et al. (2009), Marino et al. (2011), and
Villanova et al. (2012) on a few HB stars in NGC6752 and M4. The results from these 
abundance analyses agreed in general
with the hypothesis that the spread along the HB in these clusters is mostly
determined by variations of He abundances among the multiple populations present in these
clusters. On the other hand, other studies based on pulsational properties
of RR Lyrae variables, for instance, show that the situation might be more complex because at least
for some clusters, such as M3, there seems to be no evidence for variation of He, at least
close to the instability strip (see discussion in Catelan et al. 2009 and references therein). 
To further enlarge the small sample of clusters with an
extensive abundance analysis of HB stars, we started a small survey of seven GCs; 
results for five of them have already been published: NGC~2808 (Gratton et al. 2011), 
NGC~1851 (Gratton et al. 2012b), 47~Tuc and M5 (Gratton et al. 2013), and M22 (Gratton et al.
2014). In this paper we present the results for a sixth cluster, NGC~6723, while those for
the last one (NGC~6388) will be presented in a future paper. In the meantime, other studies
on this same subject appeared and included the analysis of HB stars in M~22 (Marino et al. 2013)
and NGC~6397 (Lovisi et al. 2012). The pattern emerging from all these studies generally
confirms the scenario of the multiple populations, although at least for M5
some additional mechanism should be considered.

NGC~6723 was selected for this analysis because it has an HB quite extended in colour,
ranging from stars redder than the RR~Lyrae instability strip to stars bluer than the
Grundahl jump, similar to M~5. The large extension in colours over a range where 
the abundance analysis may still provide useful results makes it a suitable cluster on which to test 
the physical mechanism causing the spread in mass of HB stars. The overall properties of NGC~6723
can clearly be estimated from accurate HST photometric studies because this cluster was 
included both in the snapshot survey by Piotto et al. (2002) and in the more recent ACS survey 
by Sarajedini et al. (2007). The papers by Sollima et al. (2007) and Milone et al. (2012), 
also using HST, were dedicated
to a search for binaries in the cluster core. Unfortunately, these photometric studies either 
focused on the central part of the cluster or on the main-sequence stars; hence there is 
little overlap with the HB stars that we observed in our spectroscopic study. 
Recent ground-based photometric studies have been performed by one of the present co-authors (PBS) and
by Lee et al. (2014). We extensively used these photometric data in our analysis. 
NGC~6723 has a low reddening (E(B-V)=0.05), a low concentration ($c=1.11$),
and a reasonably small distance modulus ($(m-M)_V=14.84$), which all facilitate
the analysis. Its overall luminosity ($M_V=-7.84$; all these values are taken from 
the compilation of data for GCs made by Harris, 1996, as downloaded from the
internet in February 2014) is somewhat above the average 
for galactic GCs. Like M~4 and M~5, NGC~6723 is a moderately metal-rich cluster: 
determinations of its metallicity include values of [Fe/H]=$-1.26\pm 0.09$\ from high-dispersion 
spectroscopy of three red giants by Fullton \& Carney (1996), $-1.14$\ from the colour-magnitude 
diagram and integrated spectrophotometry (Zinn \& West 1984), $-1.09$\ by DDO photometry (Smith \& 
Hesser 1986), $-1.35$\ from Washington photometry (Geisler 1986), 
[Fe/H]=$-1.23\pm 0.11$\ from a Fourier analysis of the light curves of RR Lyrae variables (Lee 
et al. 2014), and $-1.09\pm 0.14$\ from the Ca triplet (Rutledge et al. 1997). 
Note that almost all these values are lower than the values of [Fe/H]=$-1.10$\ 
quoted by Harris (1996, 2010 edition). None of these studies was aimed at studying multiple populations in this 
cluster. NGC~6723 is a very old cluster according to several age determinations (Mar\'in-Franch 
et al. 2009; Dotter et al. 2010; Vandenberg et al. 2013); in this respect, NGC~6723 is similar to M~4 and it looks 
older than M~5. The mean period of the RR Lyrae stars places this cluster in the Oosterhoff I group 
(see Lee et al. 2014). Finally, the orbit of NGC~6723 makes it a member of the inner halo (Dinescu 
et al. 2003), although it is projected quite close to the Galactic bulge. On the whole, NGC~6723 
appears to be a quite typical moderately metal-rich GC, its main peculiarity being the low concentration.
Unfortunately, no previous extensive study of Na and O abundances along the RGB
exist for this cluster. The anomalous extension of the HB of NGC~6723 with respect
to that of NGC~6171, for example, which has a similar metallicity, was noticed several decades ago (see 
Smith \& Hesser 1986). Gratton et al. (2010) briefly discussed this cluster: they showed that
the colour distribution along the HB looks bimodal, and it cannot be reproduced by a
single Gaussian distribution in mass. At the time, this was considered evidence for the presence of
multiple populations.

%Menzies (1974), Martins & Fraquelli
%(1987), Sarajedini (1994), Alvarado etal. (1994) e Alcaino et al. (1999).
%Una stima di metallicita' l'aveva anche data Smith (1981; [Fe/H]=-0.7) per
%le RR Lyrae e Moheler et al. (1998) avevano studiato l'HB blu

Section 2 presents the observations and data reduction, explaining how they
differ from what has been done for the other GCs studied in this series, and briefly discusses
radial and rotational velocities determined from our spectra. We derive
the atmospheric parameters and provide details of the abundance analysis in
Sect. 3. Section 4 presents the results of this analysis. Finally, in Sect. 5 we 
compare the composition of NGC~6723 with that of other GCs and compare the distribution of 
stars along the HB with that of synthetic populations specifically constructed for this purpose.

\begin{center}
\begin{figure}
\includegraphics[width=8.8cm]{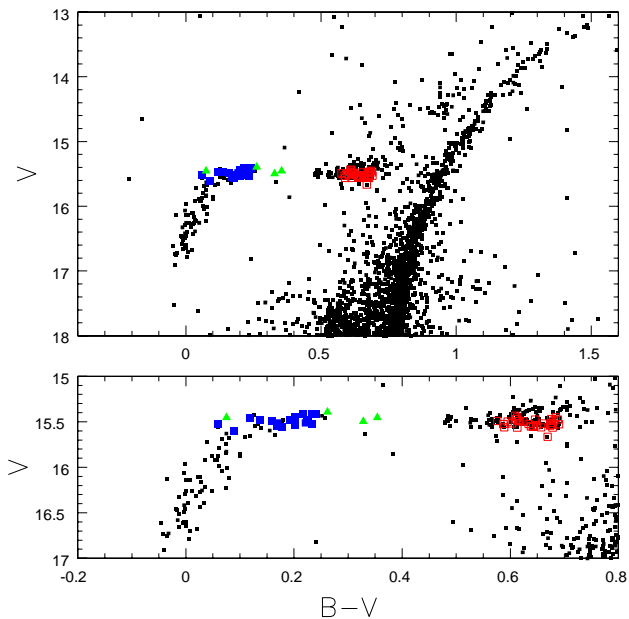}
\caption{Upper panel: ($V, B-V$) colour-magnitude diagram of NGC~6723 from Lee et al. (2014). 
lower panel: blow-up of the HB region. Stars selected
for observations are marked with large symbols: filled blue squares are BHB stars, open red
squares are RHB stars, and filled red triangles are RR Lyrae variables. }
\label{f:fig1}
\end{figure}
\end{center}

\begin{table}[htb]\centering
\caption[]{Observing log}
\setlength{\tabcolsep}{1.5mm}
\begin{tabular}{ccccccc}
\hline
Set up  &    Date     &  UT      &  Exp. time & Airmass & Seeing \\
HR      &             &  Start      & (s) & & (arcsec) \\
\hline             
12 & 2011-07-11 & 07:14:23.5 & 2800 & 1.263 & 2.18 \\
12 & 2011-08-27 & 00:10:09.5 & 2800 & 1.061 & 1.32 \\
19 & 2011-07-18 & 06:03:41.7 & 2800 & 1.145 & 0.80 \\
19 & 2011-07-18 & 06:52:18.1 & 2800 & 1.282 & 0.58 \\
\hline
\end{tabular}
\label{t:tab0}
\end{table} 

\begin{table*}[htb]
\centering
\caption[]{Photometric data (only available in electronic form)}
\begin{scriptsize}
\begin{tabular}{lccccccccc}
\hline
Star & RA (J2000) & Dec (J2000) &	V(Lee) & B(Lee) & V & B & J & H & K \\
\hline
\multicolumn{10}{c}{Blue hborizontal ranch}\\
36201 & 284.82925 & -36.67456 & 15.523 & 15.583 & 15.508 & 15.515 & 15.293 & 15.282 & 15.655 \\
36473 & 284.83075 & -36.62700 & 15.491 & 15.651 & 15.459 & 15.553 & 15.048 & 14.885 & 14.900 \\
37378 & 284.83563 & -36.60417 & 15.411 & 15.652 & 15.401 & 15.618 & 14.703 & 14.589 & 14.571 \\
44646 & 284.86296 & -36.63647 & 15.512 & 15.682 & 15.483 & 15.557 &        &        &        \\
45828 & 284.86642 & -36.65939 & 15.513 & 15.734 & 15.419 & 15.552 & 14.840 & 14.817 & 14.834 \\
47863 & 284.87225 & -36.59558 & 15.464 & 15.582 & 15.444 & 15.509 & 15.047 & 15.027 & 14.997 \\
48268 & 284.87367 & -36.61222 & 15.443 & 15.644 & 15.410 & 15.548 & 14.707 & 14.641 & 14.250 \\
49440 & 284.87775 & -36.61411 & 15.415 & 15.631 & 15.387 & 15.551 &        &        &        \\
54070 & 284.89171 & -36.65372 & 15.485 & 15.708 & 15.440 & 15.628 & 13.853 &        &        \\
54670 & 284.89321 & -36.65736 & 15.534 & 15.735 & 15.488 & 15.642 & 14.719 & 14.874 & 14.572 \\
55259 & 284.89458 & -36.66253 & 15.527 & 15.760 & 15.471 & 15.664 & 14.808 & 14.793 & 14.625 \\
56422 & 284.89750 & -36.67281 & 15.603 & 15.692 & 15.575 & 15.639 & 15.344 & 15.572 & 15.210 \\
61280 & 284.91042 & -36.60397 & 15.410 & 15.643 & 15.391 & 15.613 &        &        &        \\
63763 & 284.91788 & -36.62508 & 15.476 & 15.614 & 15.447 & 15.559 & 14.936 & 14.796 & 15.111 \\
68779 & 284.93767 & -36.62525 & 15.557 & 15.735 & 15.531 & 15.694 & 14.976 & 14.899 & 14.869 \\
71676 & 284.95396 & -36.62119 & 15.482 & 15.675 & 15.449 & 15.646 & 14.817 & 14.549 & 14.569 \\
77469 & 285.00338 & -36.55194 & 15.545 & 15.714 & 15.538 & 15.684 & 15.133 & 15.006 & 14.829 \\
\multicolumn{10}{c}{Red horizontal branch}\\
32170 & 284.80267 & -36.62553 & 15.476 & 16.164 & 15.474 & 16.141 & 14.040 & 13.594 & 13.510 \\
34102 & 284.81654 & -36.62303 & 15.470 & 16.146 & 15.467 & 16.116 & 13.972 & 13.617 & 13.535 \\
34668 & 284.82046 & -36.59639 & 15.517 & 16.120 & 15.526 & 16.122 & 14.148 & 13.795 & 13.682 \\
41649 & 284.85346 & -36.61975 & 15.492 & 16.107 & 15.473 & 16.059 & 14.104 & 13.737 & 13.677 \\
42553 & 284.85642 & -36.56922 & 15.472 & 16.083 & 15.488 & 16.100 & 14.006 & 13.659 & 13.720 \\
42857 & 284.85738 & -36.57339 & 15.435 & 16.044 & 15.447 & 16.055 & 14.114 & 13.749 & 13.609 \\
43379 & 284.85917 & -36.61406 & 15.491 & 16.108 & 15.479 & 16.067 & 14.116 & 13.691 & 13.734 \\
44426 & 284.86242 & -36.66694 & 15.665 & 16.334 & 15.530 & 16.155 & 14.071 & 13.646 & 13.574 \\
46519 & 284.86829 & -36.63739 & 15.476 & 16.122 & 15.456 & 16.069 & 13.904 &        &        \\
48528 & 284.87467 & -36.66606 & 15.525 & 16.159 & 15.508 & 16.122 & 14.064 & 13.684 & 13.653 \\
49317 & 284.87729 & -36.60628 & 15.495 & 16.119 & 15.466 & 16.076 & 14.116 & 13.766 & 13.764 \\
51670 & 284.88458 & -36.60253 & 15.452 & 16.065 & 15.445 & 16.044 & 14.076 & 13.733 & 13.629 \\
52012 & 284.88575 & -36.59378 & 15.491 & 16.067 & 15.495 & 16.073 & 14.162 & 13.842 & 13.917 \\
54960 & 284.89388 & -36.64789 & 15.538 & 16.187 & 15.501 & 16.126 &        &        &        \\
56265 & 284.89708 & -36.65158 & 15.563 & 16.241 & 15.503 & 16.162 &        &        &        \\
58421 & 284.90263 & -36.67683 & 15.532 & 16.208 & 15.506 & 16.158 & 13.992 & 13.545 & 13.516 \\
58511 & 284.90275 & -36.55697 & 15.499 & 16.094 & 15.509 & 16.112 & 14.123 & 13.808 & 13.723 \\
58978 & 284.90408 & -36.61028 & 15.542 & 16.216 & 15.465 & 16.122 &        &        &        \\
61405 & 284.91083 & -36.65272 & 15.512 & 16.188 & 15.490 & 16.153 &        &        &        \\
61969 & 284.91246 & -36.64842 & 15.520 & 16.180 & 15.496 & 16.113 & 14.029 & 13.664 & 13.613 \\
63655 & 284.91750 & -36.58208 & 15.453 & 16.137 & 15.462 & 16.149 & 13.992 & 13.571 & 13.513 \\
65270 & 284.92308 & -36.62197 & 15.561 & 16.218 & 15.493 & 16.138 & 14.038 & 13.674 & 13.580 \\
65307 & 284.92325 & -36.64167 & 15.552 & 16.190 & 15.537 & 16.185 & 14.122 & 13.745 & 13.690 \\
65835 & 284.92513 & -36.59936 & 15.525 & 16.112 & 15.522 & 16.111 & 14.185 & 13.862 & 13.661 \\
66610 & 284.92804 & -36.60747 & 15.490 & 16.170 & 15.477 & 16.156 & 14.040 & 13.651 & 13.499 \\
67352 & 284.93121 & -36.64028 & 15.563 & 16.176 & 15.544 & 16.165 & 14.166 & 13.821 & 13.764 \\
71268 & 284.95133 & -36.61556 & 15.547 & 16.189 & 15.535 & 16.179 & 14.119 & 13.754 & 13.633 \\
72060 & 284.95654 & -36.61092 & 15.561 & 16.150 & 15.560 & 16.155 & 14.240 & 13.872 & 13.817 \\
75691 & 284.98592 & -36.61997 & 15.555 & 16.198 & 15.560 & 16.197 & 14.118 & 13.769 & 13.663 \\
75716 & 284.98604 & -36.59953 & 15.528 & 16.218 & 15.533 & 16.214 & 13.982 & 13.598 & 13.547 \\
\multicolumn{10}{c}{RR Lyrae}\\
45069 & 284.86417 & -36.62147 & 15.495 & 15.823 &        &        &        &        &        \\
48493 & 284.87450 & -36.65753 & 15.392 & 15.654 &        &        &        &        &        \\
59159 & 284.90458 & -36.64344 & 15.453 & 15.807 &        &        &        &        &        \\
68073 & 284.93433 & -36.56781 & 15.453 & 15.528 &        &        &        &        &        \\
\hline
\end{tabular}
\end{scriptsize}
\label{t:tab1}
\end{table*} 
 
\section{Observations}

We observed a total of 58 candidate HB stars of NGC~6723 with FLAMES + GIRAFFE at the VLT
(Pasquini et al. 2004). The instrument was used in MEDUSA mode, with fibres pointing to 
each star and several ($\sim 20$) fibres used for determining the local sky background.
The stars were selected from the $BVI$\ photometry by PBS (see Fig.~\ref{f:fig1}); 
the field coverage is 100\% complete out to a radius of 16 arcmin, and partial
to a radius of 35 arcmin from the centre of the cluster; for comparison, the farthest
member star of NGC~6723 in our spectroscopic sample is at 7.35 arcmin from the center. We selected for
observation stars on the RHB \footnote{The RHB stars selected for observation have 
$15.445<V<15.560$\ on the original photometry used for selection. There is then a bias 
against brighter RHB stars, although membership in the cluster of all these brighter stars 
is to be demonstrated. In principle, these stars might be He-rich, but if they are He-rich,
they should also
be massive to be on the RHB. Most likely, these stars are 
simply He-normal stars evolved off the ZAHB, as shown by the simulations presented
below.} and on the BHB of the cluster, avoiding those with 
$V>15.6$\ because these stars are probably warmer than the Grundahl jump. We also tried 
to avoid placing fibres on stars within the instability strip because our observations (scheduled in 
service mode) were not optimized for variable stars. However, as reported below, four 
RR Lyrae stars were actually observed because they were erroneously recorded out of the instability 
strip in the photometry by PBS, which used observations made at only a few epochs. Only 
uncrowded stars were selected, that is, those 
without any contaminant with a $V$\ magnitude brighter than $V_*+2$\ within 2~arcsec from 
it; here $V_*$\ is the magnitude of the programme star,. For this reason and to
avoid fiber-to-fiber collisions, the star closest to the cluster center is at 0.98 arcmin.
Seven of the observed stars (all close 
to the RHB of the cluster in the colour-magnitude diagram) were not members 
of NGC~6723 because they have discrepant radial velocities and abundances. These non-members 
were typically projected at large distances from the 
cluster center, but still within the tidal radius. Of the remaining 51 bona-fide members of NGC~6723, 17 are BHB stars, 
30 are RHB stars, and the remaining four are RR Lyrae variables according to
cross-identification with the list by Lee et al. (2014). The UVES fibres were used to 
acquire spectra of stars on the asymptotic giant branch of the cluster; they will be 
analysed in a future paper.

As was the case for the observations of the other GCs of this series, we used two set-ups: HR12
(wavelength range 5821-6146~\AA, resolution $R\sim 18,700$), and HR19A (wavelength range
7745-8335~\AA, resolution $R\sim 13,867$), allowing
observation of the Na~D doublet and the O~I triplet at 7771-74~\AA. These are the strongest
features due to Na and O in the spectra of HB stars that are accessible from the ground.
We aimed to obtain a final S/N=50 per pixel. The observations (see Table~\ref{t:tab0}) were carried out in 
service mode. The first frame with the HR12 set-up was
acquired in poor observing conditions, yielding spectra with a S/N markedly lower than
the other exposures (see Col. 2-5 of Table~\ref{t:tab2}). We used these data only to 
derive radial velocities (see Col. 6-9 of the same table), the abundance analysis 
is based on the remaining spectra.

Photometric data for the programme stars are listed in Table~\ref{t:tab1}. It includes the
$BV$\ photometry by PBS (in preparation), as well as the high quality $BV$\ photometry by
Lee et al. (2014), kindly provided by the first author of that paper. In addition,
we also listed the $JHK_S$\ photometric data from the 2MASS catalogue (Skrutskie et al.
2006). Since the observed stars are not in the inner region of the cluster, the WFPC2
photometry of Piotto et al. (2002) is of little help. We also searched for stars in 
common with the ACS photometry by Sarajedini et al. (2007), but we could not 
cross-identify any star because the observed stars are brighter than those listed in that catalogue.

Spectra were reduced using the ESO GIRAFFE pipeline, which provides wavelength-calibrated 
spectra. Sky subtraction was performed using IRAF \footnote{IRAF is distributed by the 
National Optical Astronomy Observatory, which is operated by the Association of 
Universities for Research in Astronomy (AURA) under cooperative agreement with the 
National Science Foundation}.
We selected not to sum the individual spectra, but rather to measure velocities and
equivalent widths on each of them, and then averaged the results of the analysis.
The values from individual spectra are labelled as A and B in Table~\ref{t:tab2}.

\begin{center}
\begin{figure}
\includegraphics[width=8.8cm]{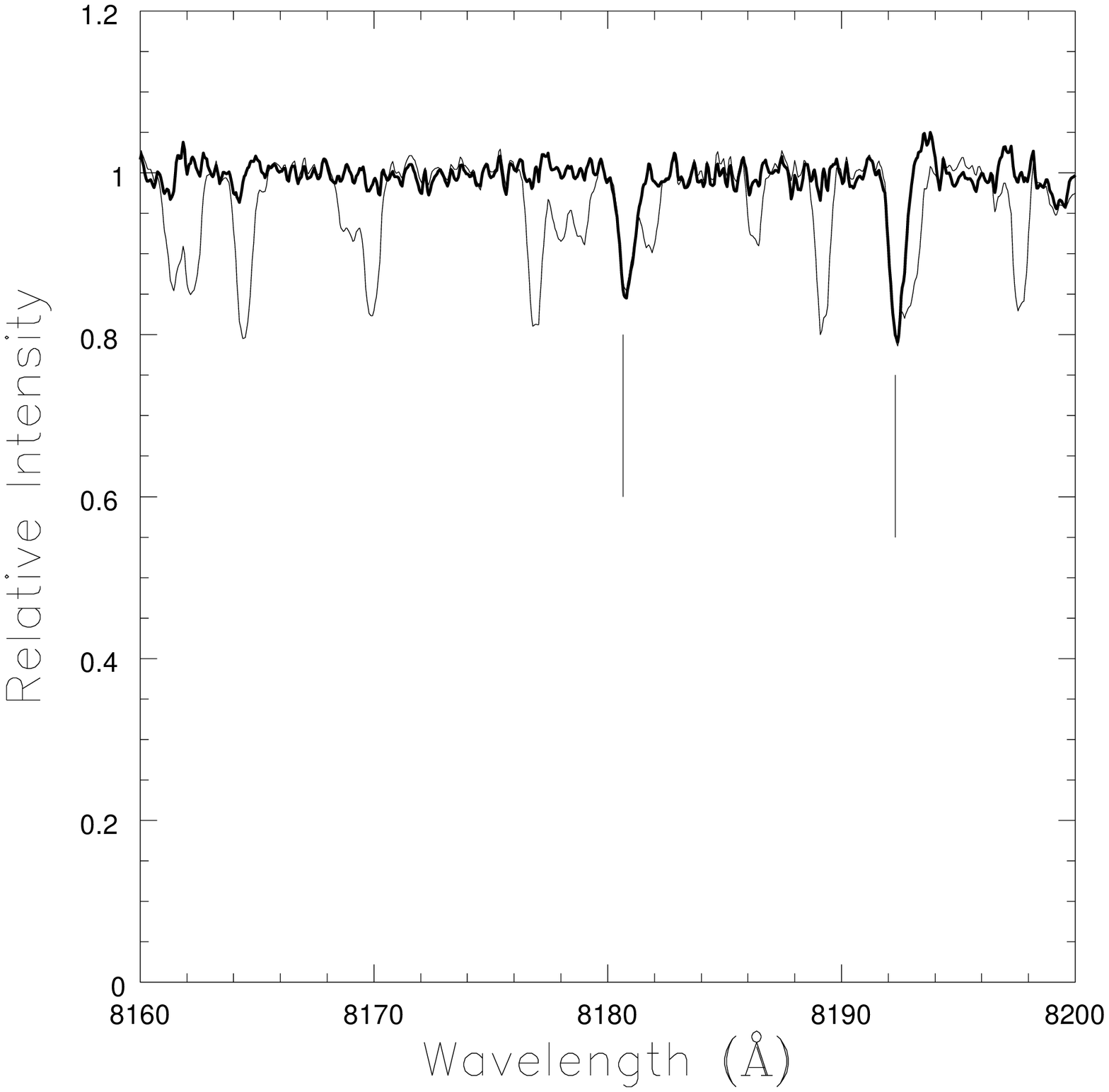}
\caption{Example of subtraction of telluric lines from the spectra obtained with the HR19A
set-up following the procedure described in the text (thin line: before division;
thick line: after division). The marks indicate the Na doublet
at 8183-94~\AA, slightly shifted blueward due to the geocentric radial velocity of the
star (\#32170) at the epoch of observation.}
\label{f:figsub}
\end{figure}
\end{center}

Telluric absorption lines were removed from the HR19A spectra by dividing them by
an average spectrum of a few extreme BHB stars in NGC~1851 (Gratton et al. 2012b).
Figure~\ref{f:figsub} shows an example of such a subtraction. We deem this subtraction
to be fully satisfactory.

\begin{table*}[htb]
\centering
\caption[]{S/N of spectra, radial velocities, FWHM, and rotational velocities (only available in electronic form)}
\setlength{\tabcolsep}{1.5mm}
\begin{scriptsize}
\begin{tabular}{lcccccccccccccc}
\hline
Star&S/N(12A)&S/N(12B)&S/N(19)&S/N(19B)&RV(12A)&RV(12B)&RV(19A)&RV(19B)&$<$RV$>$&r.m.s. &$<$FWHM$>$& r.m.s. &$v\sin{i}$ & rms \\
    &        &        &  &  &(km~s$^{-1}$)&(km~s$^{-1}$)&(km~s$^{-1}$)&(km~s$^{-1}$)&(km~s$^{-1}$)&(\AA)&(\AA)&(km~s$^{-1}$)&(km~s$^{-1}$)\\
\hline
\multicolumn{15}{c}{Blue horizontal branch}\\
36201 & 16 & 40 & 39 & 36 & -97.19 & -96.81 & -96.10 & -93.67 & -95.94 & 1.58 & 0.76 & 0.04 & 10.4 & 2.2 \\
36473 & 21 & 40 & 52 & 53 & -92.08 & -90.52 & -89.96 & -90.32 & -90.72 & 0.94 & 0.69 & 0.02 &  5.1 & 2.1 \\
37378 & 21 & 41 & 53 & 50 &-100.57 & -99.13 & -99.43 & -97.53 & -99.17 & 1.25 & 0.73 & 0.03 &  8.8 & 1.9 \\
44646 & 22 & 30 & 38 & 47 & -98.73 & -96.94 &-100.13 &-100.72 & -99.13 & 1.68 & 0.76 & 0.02 & 10.2 & 0.8 \\
45828 & 22 & 39 & 52 & 42 & -92.38 & -92.67 & -90.58 & -92.23 & -91.97 & 0.94 & 0.67 & 0.03 &$<$5.0&     \\
47863 & 24 & 32 & 62 & 43 &-102.49 &-101.54 &-102.48 &-100.93 &-101.86 & 0.76 & 0.65 & 0.03 &$<$5.0&     \\
48268 & 25 & 37 & 63 & 43 & -92.26 & -91.82 & -90.63 & -92.72 & -91.86 & 0.89 & 0.69 & 0.03 &  5.4 & 2.4 \\
49440 & 24 & 33 & 53 & 43 & -96.63 & -94.99 & -95.50 & -95.58 & -95.67 & 0.69 & 0.73 & 0.01 &  8.8 & 0.4 \\
54070 & 23 & 40 & 46 & 41 & -89.98 & -90.68 & -91.21 & -91.06 & -90.73 & 0.55 & 0.84 & 0.04 & 14.3 & 1.8 \\
54670 & 14 & 46 & 42 & 48 & -93.28 & -92.22 & -90.66 & -91.73 & -91.97 & 1.09 & 0.73 & 0.05 &  8.4 & 2.7 \\
55259 & 21 & 36 & 55 & 48 & -95.11 & -92.78 & -92.85 & -93.88 & -93.66 & 1.09 & 0.66 & 0.03 &$<$5.0&     \\
56422 & 20 & 36 & 41 & 49 & -88.88 & -89.84 & -88.40 & -87.84 & -88.74 & 0.85 & 0.72 & 0.04 &  8.1 & 2.6 \\
61280 & 28 & 31 & 45 & 52 & -95.57 & -95.76 & -94.81 & -94.36 & -95.13 & 0.65 & 0.69 & 0.06 &  5.9 & 3.9 \\
63763 & 18 & 40 & 33 & 35 & -97.51 & -95.21 & -94.63 & -97.30 & -96.16 & 1.46 & 0.66 & 0.02 &$<$5.0&     \\
68779 & 20 & 36 & 42 & 40 & -95.38 & -92.93 & -93.20 & -96.61 & -94.53 & 1.77 & 0.64 & 0.05 &$<$5.0&     \\
71676 & 26 & 34 & 46 & 38 & -94.35 & -93.67 & -92.13 & -95.36 & -93.88 & 1.36 & 0.66 & 0.04 &$<$5.0&     \\
77469 & 26 & 30 & 42 & 40 & -95.82 & -96.21 & -95.94 & -95.93 & -95.98 & 0.17 & 0.64 & 0.03 &$<$5.0&     \\
\multicolumn{15}{c}{Red horizontal branch}\\
32170 & 21 & 34 & 55 & 54 & -95.32 & -94.81 & -92.95 & -96.42 & -94.88 & 1.45 &      &      &      &     \\				
34102 & 21 & 34 & 50 & 62 &-100.94 &-100.29 &-101.36 &-102.32 &-101.23 & 0.85 &      &      &      &     \\				
34668 & 23 & 23 & 59 & 54 & -94.60 & -94.88 & -93.56 & -92.73 & -93.94 & 0.99 &      &      &      &     \\				
41649 & 27 & 33 & 57 & 62 & -96.71 & -95.93 & -95.23 & -95.82 & -95.92 & 0.61 &      &      &      &     \\				
42553 & 20 &    & 43 & 44 &-102.77 &        &-100.32 & -99.38 &-100.83 & 1.75 &      &      &      &     \\				
42857 & 22 & 35 & 53 & 57 & -95.64 & -95.18 & -93.16 & -94.40 & -94.60 & 1.09 &      &      &      &     \\				
43379 & 22 & 41 & 34 & 36 & -98.07 & -97.90 & -98.23 & -96.34 & -97.63 & 0.87 &      &      &      &     \\				
44426 & 18 & 36 & 63 & 50 &-105.16 &-104.73 &-104.65 &-104.12 &-104.67 & 0.43 &      &      &      &     \\				
46519 & 28 & 48 & 35 & 40 & -86.41 & -84.23 & -82.68 & -81.90 & -83.80 & 1.99 &      &      &      &     \\				
48528 & 22 & 42 & 46 & 39 & -94.86 & -95.06 & -92.48 & -94.79 & -94.30 & 1.22 &      &      &      &     \\				
49317 & 25 & 33 & 52 & 50 & -96.01 & -94.97 & -93.95 & -96.19 & -95.28 & 1.04 &      &      &      &     \\				
51670 & 15 & 33 & 37 & 51 & -99.88 & -98.68 & -98.24 &-101.39 & -99.55 & 1.41 &      &      &      &     \\				
52012 & 22 & 36 & 75 & 48 &-105.32 &-104.77 &-103.39 &-103.44 &-104.23 & 0.97 &      &      &      &     \\				
54960 & 31 & 39 & 56 & 51 & -93.69 & -92.94 & -92.14 & -94.98 & -93.44 & 1.21 &      &      &      &     \\				
56265 & 26 & 33 & 68 & 61 & -99.95 & -99.82 & -99.53 &-101.21 &-100.13 & 0.74 &      &      &      &     \\				
58421 & 26 & 39 & 55 & 52 & -97.37 & -96.18 & -98.75 & -98.53 & -97.71 & 1.19 &      &      &      &     \\				
58511 & 21 & 38 & 61 & 49 & -93.25 & -92.84 & -91.77 & -91.42 & -92.32 & 0.87 &      &      &      &     \\				
58978 & 29 & 35 & 37 & 38 & -89.61 & -88.77 & -85.37 & -87.76 & -87.88 & 1.83 &      &      &      &     \\				
61405 & 27 & 30 & 48 & 50 & -96.18 & -95.57 & -96.51 & -97.52 & -96.45 & 0.82 &      &      &      &     \\				
61969 & 28 & 39 & 50 & 69 &-104.40 &-103.55 &-104.46 &-104.41 &-104.21 & 0.44 &      &      &      &     \\				
63655 & 23 & 44 & 45 & 51 & -96.84 & -96.07 & -95.51 & -98.79 & -96.80 & 1.43 &      &      &      &     \\				
65270 & 24 & 30 & 58 & 62 & -99.88 & -98.54 &-101.84 & -96.23 & -99.12 & 2.36 &      &      &      &     \\				
65307 & 22 & 39 & 45 & 50 & -94.87 & -93.99 & -96.63 & -94.53 & -95.00 & 1.15 &      &      &      &     \\				
65835 & 22 & 31 & 53 & 39 & -91.06 & -90.76 & -91.68 & -91.53 & -91.26 & 0.42 &      &      &      &     \\				
66610 & 23 & 36 & 59 & 51 & -98.40 & -96.59 & -99.14 & -99.28 & -98.35 & 1.23 &      &      &      &     \\				
67352 & 22 & 36 & 62 & 40 &-100.16 & -99.88 &-100.12 & -99.93 &-100.02 & 0.14 &      &      &      &     \\				
71268 & 24 & 46 & 45 & 41 & -94.83 & -94.55 & -94.42 & -93.67 & -94.37 & 0.49 &      &      &      &     \\				
72060 & 23 & 34 & 51 & 38 & -91.45 & -90.69 & -87.47 & -91.06 & -90.17 & 1.83 &      &      &      &     \\				
75691 & 20 & 34 & 47 & 46 & -95.06 & -94.15 & -96.08 & -96.05 & -95.34 & 0.92 &      &      &      &     \\				
75716 & 22 & 36 & 44 & 52 & -99.91 & -98.48 &-100.04 &-101.62 &-100.01 & 1.28 &      &      &      &     \\				
\multicolumn{15}{c}{RR Lyrae}\\
45069 & 26 & 31 & 49 & 47 &-109.30 & -94.57 & -87.04 & -88.54 & -94.86 &10.16 &      &      &      &     \\				
48493 & 25 & 41 & 64 & 41 &-100.33 & -93.77 &-100.85 &-105.47 &-100.10 & 4.82 &      &      &      &     \\				
59159 & 23 & 29 & 42 & 43 & -68.49 & -81.44 & -69.96 & -70.15 & -72.51 & 6.00 &      &      &      &     \\				
68073 & 27 & 34 & 43 & 48 &        &        & -67.37 & -66.44 & -66.90 & 0.66 &      &      &      &     \\
\hline
\end{tabular}
\end{scriptsize}
\label{t:tab2}
\end{table*} 

%\begin{center}
%\begin{figure}
%\includegraphics[width=8.8cm]{M22-vrads.ps}
%\caption{Radial velocities from set-up HR03 vs those from set-up HR19A; 
%different colours are for stars of different groups (see Section 4): Group 1: 
%red filled squares; Group 2: black open squares; Group 3: blue filled triangles.
%Dotted line represents equality; dashed lines are $\pm 2$~times the observational
%errors.}
%\label{f:fig2}
%\end{figure}
%\end{center}
                  
\subsection{Radial velocities}

As mentioned above, radial velocities were measured on the individual spectra using a
few selected lines. Average radial velocities and the r.m.s. scatter around them are 
given in Col. 10 and 11 of Table~\ref{t:tab2}. The average value over 
all stars (excluding the RR~Lyrae variables) is $-95.8\pm 0.6$~km~s$^{-1}$, with an 
r.m.s. scatter for individual stars of 4.4~km~s$^{-1}$. The median value of the r.m.s. 
scatter from different measurements for an individual star is about 1~km~s$^{-1}$, which 
is much smaller than the star-to-star scatter. For comparison, Harris (1996, 2010 edition) listed an 
average velocity for NGC~6723 of $-94.5\pm 3.6$~km~s$^{-1}$, in good agreement with our 
determination, considering the large errors and scatter of the individual values from which
this result is obtained (Zinn \& West 1984: $-90\pm 20$~km~s$^{-1}$; Hesser et al. 1986: 
$-79\pm 7$~km~s$^{-1}$; Rutledge et al. 1997: $-100.3\pm 9.9$~km~s$^{-1}$). The r.m.s. 
scatter we obtained implies an internal velocity dispersion of 4.3 km/s, which represents the first
measure of this quantity in NGC~6723.
We note here that the stars we observed are at a largest/smallest projected distance
from the cluster center of 0.98/7.35 arcmin, with a median value of 2.3 arcmin. For comparison,
the core, tidal, and half-light radii of NGC~6723 are 0.83, 13.7, and 1.53 arcmin,
respectively (data from Harris 1996). Our data then refer to the outer regions of the
cluster. 

We found a small systematic difference when we considered BHB and RHB stars separately.
The average radial velocities are $-94.5\pm 0.8$~km~s$^{-1}$\ for BHB and $-96.4\pm 0.9$~km~s$^{-1}$\
RHB stars. The difference ($1.9\pm 1.2$~km~s$^{-1}$) is only 
marginally significant. If real, it might indicate a highest convective blue-shift for the 
RHB stars with respect to the BHB. Trying to better establish this point, we considered 
all those clusters for which we have radial velocities from both BHB and RHB stars from 
this survey. In addition to NGC~6723, we have data for three other clusters: NGC1851,
NGC2808, and M~5, for which we found radial velocity differences of $1.9\pm 0.9$, 
$7.8\pm 4.0$, and $1.4\pm 1.0$~km~s$^{-1}$, respectively. In all cases, the offset is in the sense
that the radial velocities for BHB stars are higher than those for RHB. The weighted
average of all these data is $1.9\pm 0.6$~km~s$^{-1}$. The offset is significant at more
than 3$\sigma$. Since we expect no real systematic difference in the average velocities of
BHB and RHB stars, we assume that this effect is due to systematic errors in the measurements.
Given that the same lines are used for the two sets of stars, we suggest that RHB 
stars have a convective blue-shift higher than the BHB stars (note that Na D lines have 
quite a strong weight in our radial velocities). Regardless of its cause, this effect should 
be considered before combining data from stars in these evolutionary phases for dynamical 
analysis of the clusters.

Finally, there is no strong indication favouring intrinsic variability of the radial
velocities for any of the non-variable stars we observed, although 
our data are certainly not ideal for detecting spectroscopic binaries. On the other
hand, quite large variations were obtained for three of the four RR Lyrae variables we observed.

\begin{table*}[htb]
\centering
\caption[]{Fraction of fast rotators ($V\sin{i}>20$~km~s$^{-1}$) among BHB stars  cooler than
the Grundahl jump}
\setlength{\tabcolsep}{1.5mm}
\begin{tabular}{lcccccl}
\hline
NGC & Messier & $[$Fe/H$]$ & HBR & N(BHB) & N(Fast) & Source \\
\hline
6341 & M92 & -2.35 & 0.91 & 22 &  7 & Behr 2003a \\
7078 & M15 & -2.33 & 0.67 & 25 &  3 & Recio-Blanco et al. 2004 \\
     &     &       &      &    &    & Behr et al. 2000b \\
     &     &       &      &    &    & Behr 2003a \\    
4590 & M68 & -2.27 & 0.17 & 11 &  3 & Behr 2003a \\
6397 &     & -1.98 & 0.98 & 40 & 23 & Lovisi et al 2012 \\
6093 & M80 & -1.75 & 0.93 & ~7 &  0 & Recio-Blanco et al. 2004 \\
6656 & M22 & -1.70 & 0.91 & 92 & 34 & Gratton et al. 2014 \\
6205 & M13 & -1.58 & 0.97 & 28 &  8 & Behr et al. 2000a \\
     &     &       &      &    &    & Behr 2003a \\
1904 & M79 & -1.58 & 0.89 & 12 &  7 & Recio-Blanco et al. 2004 \\
5272 & M3  & -1.50 & 0.08 & 28 &  3 & Behr 2003a \\
5904 & M5  & -1.33 & 0.31 & 44 &  5 & Gratton et al. 2013 \\
288	 &     & -1.32 & 0.98 & 16 &  0 & Behr 2003a \\
6723 &     & -1.22 &-0.08 & 17 &  0 & This paper \\
1851 &     & -1.18 &-0.36 & 27 &  1 & Gratton et al. 2012b \\
6121 & M4  & -1.18 &-0.06 & ~6 &  0 & Villanova et al. 2012 \\
2808 &     & -1.18 &-0.49 & ~9 &  0 & Recio-Blanco et al. 2004 \\
\hline
\end{tabular}
\label{t:tabfast}
\end{table*} 

\begin{center}
\begin{figure}
\includegraphics[width=8.8cm]{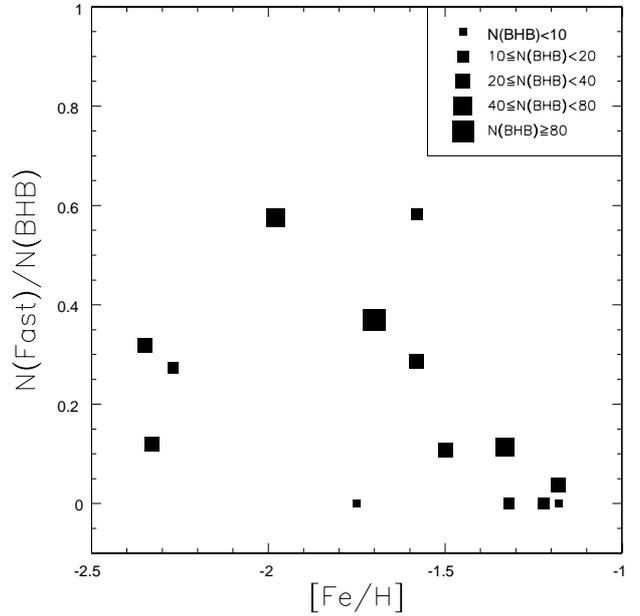}
\caption{Fraction of fast rotators ($V\sin{i}>20$~km~s$^{-1}$) among BHB stars
as a function of cluster [Fe/H] value. Symbol size represents the number of
stars that have been examined in the different clusters}
\label{f:figfast}
\end{figure}
\end{center}

\subsection{Rotational velocities}

It is well known that a fraction of the BHB stars rotate at moderate velocity (up
to several tens km~s$^{-1}$; Peterson 1983; Behr et al. 1999; Recio-Blanco et al. 2004).
Generally, fast rotators are found only among BHB stars cooler than the Grundhal jump
(Behr 2003a; Recio-Blanco et al. 2004; Lovisi et al. 2012); only a few RHB stars have been found to rotate
(Behr et al. 2003b) and most likely they result from the evolution in close binary systems 
(see the case of a fast rotator in M~5 discussed in Gratton et al. 2013). We call the stars 
with $V\sin{i}>20$~km~s$^{-1}$\ fast rotators. We may use our spectra to search for 
fast rotators among the BHB stars we are observing in NGC~6723 (none of the RHB stars 
have lines wider than expected from instrumental effects). Since other lines are weak, 
we obtained the full width at half maximum (FWHM) of the O~I spectral lines. They are given 
in Col. 12 of Table~\ref{t:tab2}, along with their uncertainties as estimated from the 
scatter of individual measurements (Col. 13). They were transformed to rotational 
velocities following the same method as described in Gratton et al. (2014). As in that paper, 
whenever a lower value was obtained, we assumed an upper limit of 5~km~s$^{-1}$.
Most stars are slow rotators ($v~\sin{i}<10$~km~s$^{-1}$); only three stars rotate at 
moderate speed, but none is a fast rotator according to the previous definition. This 
result is clearly at odds with what we recently obtained for the BHB stars in M~22, 
where we found 34 fast rotators among 92 stars (Gratton et al. 2014).

More in general, the fraction of fast rotators among BHB stars (here, only stars cooler 
than the Grundahl jump are considered) seems to be a function of the cluster 
metallicity and of the HB colour as represented by the HBR index\footnote{HBR is defined as
[n(BHB)-n(RHB)]/[n(BHB+n(RR)+n(RHB)], where n(BHB), n(RHB), n(RR) are the number of HB stars
bluer than, redder than, and within the instability strip, respectively.}. 
Table~\ref{t:tabfast} collects data available for 15 GCs. We also list [Fe/H] 
values and HBR ratios from Harris (1996, 2010 edition). These data were used to construct 
Fig.~\ref{f:figfast}, where we plot the fraction of fast rotators among BHB stars
as a function of [Fe/H]. The highest percentages of fast rotators (30-70\%) are obtained 
for moderately metal-poor clusters (NGC~6397, M~22, M~13, and M~79) that typically have 
very blue HBs. A lower percentage of between 10-30\% is obtained for the most metal-poor 
clusters (M~92, M~15, and M~68), which are typical Oosterhoff~II clusters. Fast rotators 
are rare among BHB stars in metal-rich  clusters ([Fe/H]$>-1.3$). Most of these last 
clusters are Oosterhoff~I clusters. However, fast rotators are also absent from the 
BHB stars of the metal-rich cluster NGC~288, and none have been found in the 
metal-poor cluster NGC~6093 (however, data are available for only seven BHB stars in 
this cluster). Both NGC~288 and NGC~6093 have very blue HBs. There is then a clear connection,
if not a one-to-one correlation, between rotation and both metallicity and colours 
along the HB. However, to clarify the meaning of this connection, an explanation for 
the lack of fast rotators among stars warmer than the Grundahl jump is required
(Behr 2003a; Recio-Blanco et al. 2004). For possible suggestions, see Sweigart (2002)
and Vink and Cassisi (2002).

%: these results suggest
%that surface rotation is modulated by the size of the outer convective envelope of
%HB stars.

\begin{table}[htb]
\centering
\caption[]{Atmospheric parameters (only available in electronic form)}
\begin{scriptsize}
\begin{tabular}{lcccccccccc}
\hline
Star& $T_{\rm eff}$& Err & $\log{g}$ & $v_t$ & $[$A/H$]$ \\
    & (K)& (K)& (dex) & (km~s$^{-1}$)& (dex)\\
\hline
\multicolumn{6}{c}{Blue horizontal branch}\\
36201 & 9312 & 448 & 3.32 & 3.00 & -1.25 \\
36473 & 8308 & 219 & 3.17 & 3.42 & -1.25 \\
37378 & 7711 &  57 & 3.03 & 3.77 & -1.25 \\
44646 & 8070 & 490 & 3.13 & 3.56 & -1.25 \\
45828 & 7997 &  48 & 3.09 & 3.60 & -1.25 \\
47863 & 8652 & 259 & 3.21 & 3.21 & -1.25 \\
48268 & 8037 &  59 & 3.10 & 3.58 & -1.25 \\
49440 & 7930 &  12 & 3.07 & 3.64 & -1.25 \\
54070 & 7849 &  44 & 3.07 & 3.69 & -1.25 \\
54670 & 7995 &  26 & 3.12 & 3.60 & -1.25 \\
55259 & 7807 &  56 & 3.08 & 3.72 & -1.25 \\
56422 & 8847 & 383 & 3.29 & 3.09 & -1.25 \\
61280 & 7721 &  10 & 3.03 & 3.77 & -1.25 \\
63763 & 8324 &  65 & 3.16 & 3.41 & -1.25 \\
68779 & 8032 &  67 & 3.15 & 3.58 & -1.25 \\
71676 & 7904 &  46 & 3.09 & 2.00 & -1.25 \\
77469 & 8104 & 219 & 3.16 & 3.54 & -1.25 \\
\multicolumn{6}{c}{Red horizontal branch}\\
32170 & 5482 &  75 & 2.47 & 1.30 & -1.25 \\	% 8.2
34102 & 5499 &  33 & 2.48 & 1.30 & -1.25 \\	%--
34668 & 5661 &  50 & 2.56 & 1.30 & -1.25 \\	% 7.9
41649 & 5661 &  51 & 2.54 & 1.30 & -1.25 \\	%--
42553 & 5611 &  69 & 2.53 & 1.30 & -1.25 \\	%11.3
42857 & 5651 &  25 & 2.52 & 1.30 & -1.25 \\	%--
43379 & 5657 &  46 & 2.54 & 1.30 & -1.25 \\	%12.6
44426 & 5540 &  50 & 2.52 & 1.30 & -1.25 \\	% 9.7 xxx
46519 & 5556 &  67 & 2.49 & 1.30 & -1.25 \\	%--
48528 & 5592 &  29 & 2.53 & 1.30 & -1.25 \\	
49317 & 5628 &   5 & 2.52 & 1.30 & -1.25 \\	% 3.6
51670 & 5648 &  33 & 2.52 & 1.30 & -1.25 \\	%--
52012 & 5722 &  81 & 2.57 & 1.30 & -1.25 \\	
54960 & 5565 &  25 & 2.51 & 1.30 & -1.25 \\	%11.4
56265 & 5466 &  13 & 2.48 & 1.30 & -1.25 \\ %13.7	
58421 & 5492 &  26 & 2.49 & 1.30 & -1.25 \\	
58511 & 5660 &  64 & 2.55 & 1.30 & -1.25 \\	%--
58978 & 5476 &   9 & 2.47 & 1.30 & -1.25 \\	%12.2
61405 & 5463 &   0 & 2.47 & 1.30 & -1.25 \\	%--
61969 & 5556 &  49 & 2.51 & 1.30 & -1.25 \\	%--
63655 & 5458 &  78 & 2.46 & 1.30 & -1.25 \\	%10.0
65270 & 5532 &  19 & 2.50 & 1.30 & -1.25 \\	% 9.1
65307 & 5558 &  41 & 2.53 & 1.30 & -1.25 \\	% 9.4
65835 & 5696 &  62 & 2.57 & 1.30 & -1.25 \\	% 7.9
66610 & 5477 &  79 & 2.47 & 1.30 & -1.25 \\	%--
67352 & 5622 &  38 & 2.55 & 1.30 & -1.25 \\	%--
71268 & 5557 &  31 & 2.52 & 1.30 & -1.25 \\	%--
72060 & 5691 &  53 & 2.58 & 1.30 & -1.25 \\	%--
75691 & 5558 &  12 & 2.53 & 1.30 & -1.25 \\	%--
75716 & 5440 &  44 & 2.48 & 1.30 & -1.25 \\	%--
\hline
\end{tabular}
\end{scriptsize}
\label{t:tab3}
\end{table} 

\section{Analysis}

We analysed only the spectra of non-variable stars. Reliable abundances for the
RR Lyrae stars would have required scheduled observations taken at appropriate phases; this
was not possible for the adopted service mode for the observations.

\subsection{Atmospheric parameters}

Our analysis is based on model atmospheres extracted by interpolation within the Kurucz 
(1993) grid, with the overshooting option switched off. Interpolations were made as 
described in Gratton \& Sneden (1987) and as 
used in many other papers. The grid of models used for this interpolation does not include 
any enhancement of the $\alpha-$elements.
The effect of $\alpha-$enhancement, typically observed in metal-poor stars, is expected to be 
weak for atmospheres with $T_{\rm eff}>4500$~K (see e.g. Gustafsson et al. 2008).
The main effect is expected on the continuum opacity. For RHB stars, opacity is dominated by H${}^-$, 
with electrons mainly provided by metals. Alpha-enhanced models ([$\alpha$/Fe]=+0.4) should then 
resemble $\alpha$-normal models that are more metal-rich by 0.2-0.3~dex. Given the sensitivity to metal 
abundance reported below, the effect is minor for almost all species considered 
in our analysis. The Ba abundances might be systematically underestimated by as much as 
0.1~dex, but this effect probably is the same for all stars, so that it does not affect the 
star-to-star trends considered in this paper. $\alpha$-enhancement has an even weaker effect 
on BHB stars, since in this case opacity is mainly due to H, which also contributes most of the 
free electrons.

The most critical parameter in our abundance analysis is the effective temperature \teff. 
The values we adopted were obtained from the de-reddened $B-V$\ colours and, for RHB stars, 
also from the $V-J$\ colours ($K$\ magnitudes typically have much larger errors and we
prefer not to use them). We used the calibration by Alonso et al. (1999) for the RHB 
stars, while for the BHB stars we used the same calibration as was adopted for the stars 
in M~5 by Gratton et al. (2013). The reddening value we adopted is $E(B-V)=0.05$\ from
Harris (1996, 2010 edition). Star-to-star uncertainties in these values for \teff\ can be obtained
by comparing results from different photometric studies. The median of
the r.m.s. dispersion is 59~K for the BHB stars and 43~K for RHB stars. Systematic
errors are probably larger. We assume that they are about 100~K for RHB stars and twice
that for the BHB stars.

Surface gravities were obtained from visual magnitudes and effective temperatures, using 
the distance modulus $(m-M)_V=14.84$~mag from Harris (1996, 2010 edition), the bolometric corrections by 
Alonso et al. (1999) for the RHB stars and from Kurucz (1993) for the BHB ones, and masses of
0.610 and 0.664~M$_\odot$\ for the BHB and RHB stars. These last values
are taken from the Gratton et al. (2010) analysis of the statistical properties of the HB 
stars. While there are some uncertainties in these adopted quantities, 
the total errors in the gravities are small ($<0.1$~dex).

There are too few lines in our spectra for a reliable determination of the microturbulence
velocity. We therefore adopted a value of 1.3~km~s$^{-1}$\ for all RHB stars: this value is
intermediate between those adopted for 47~Tuc and M~5 RHB stars in Gratton et al. (2013).
For the BHB stars, we used the values given by the relations $v_t=3.0$~km~s$^{-1}$\ 
for stars with \teff$>9000$~K, and $v_t=3.0- 0.6($\teff$-9000)$\ (Gratton
et al. 2014), save for star \#71676, for which we adopted a lower value of 2.0~km~s$^{-1}$\
to reach reasonable agreement among different lines of the same element.
These values are quite uncertain, mainly for BHB stars, where we deem that the error can be as 
large as 0.5~km~s$^{-1}$\ from the scatter found between different BHB studies (compare
for instance the values of Marino et al. 2013 with those of For \& Sneden 2010). 
There is much better agreement among different analyses of RHB stars, so that the 
adopted microturbulence velocities for these stars probably have errors not larger 
than 0.2~km~s$^{-1}$.

Finally, we adopted a model metal abundance of [A/H]=$-1.25$, close to the average value we determined
for the Fe abundance.

\subsection{Equivalent widths}

Our abundances rest on measures of equivalent widths ($EW$). In most cases, they were 
obtained by an automatic procedure analogous to that used in many papers of our team on 
red giants (see, e.g., Carretta et al. 2009). The results from this procedure are not
accurate for strong lines ($EW>150$~m\AA). In these cases, the $EW$s were obtained
from manual measurements. Since we obtained independent measures from the two spectra 
obtained with the HR19A set up, we may compare the two values and derive an estimate of 
the internal error of our $EW$s. The value we obtain ($\pm 4.6$~m\AA) compares well with 
that determined using the Cayrel (1988) formula. We estimated internal errors in the 
$EW$s for the HR12 spectra from comparing results for stars with very similar atmospheric 
parameters. Since we only used one spectrum per star, the errors are larger 
($\pm 6.9$~m\AA).

\begin{table}[htb]
\centering
\caption[]{Sensitivity of abundances on the atmospheric parameters and total errors}
\setlength{\tabcolsep}{1.5mm}
\begin{tabular}{lcccccc}
\hline
Element & \teff & $\log{g}$ & $v_t$ & $[$A/H$]$ & $EW$ & Total \\
        & (K) & & (km~s$^{-1}$) & & (m\AA) & \\
\hline
Variation      &+100&+0.3&+0.5&+0.2&+10\\
\multicolumn{7}{c}{RHB star}\\
\hline
Error          &  50& 0.1 & 0.2 & 0.1 & 5 &  \\
\hline
$[$Fe/H$]$~I   & ~0.087 & -0.018 & -0.120 & -0.014 & ~0.027 & 0.07 \\ 
$[$O/Fe$]$~I   & -0.208 & ~0.126 & -0.045 & ~0.036 & ~0.087 & 0.12 \\
$[$Na/Fe$]$~I  & ~0.030 & -0.119 & ~0.047 & ~0.042 & ~0.075 & 0.06 \\
$[$Mg/Fe$]$~I  & -0.037 & -0.032 & ~0.048 & ~0.007 & ~0.150 & 0.08 \\
$[$Si/Fe$]$~I  & -0.051 & ~0.013 & ~0.067 & ~0.017 & ~0.075 & 0.05 \\
$[$Ca/Fe$]$~I  & ~0.013 & -0.052 & ~0.012 & ~0.018 & ~0.106 & 0.06 \\
$[$Ni/Fe$]$~I  & ~0.018 & ~0.021 & ~0.051 & ~0.005 & ~0.087 & 0.05 \\
$[$Ba/Fe$]$~II & -0.049 & ~0.093 & -0.311 & ~0.077 & ~0.150 & 0.16 \\
\hline
\multicolumn{7}{c}{BHB}\\
\hline
Error          & 100 & 0.1 & 0.5 & 0.2 & 5 &  \\
\hline
$[$N/Fe$]$~I   &  0.018 & ~0.049 & -0.057 & -0.001 & ~0.065 & 0.07 \\
$[$O/Fe$]$~I   &  0.021 & -0.002 & -0.154 & -0.012 & ~0.099 & 0.16 \\
$[$Na/Fe$]$~I  &  0.087 & -0.169 & -0.057 & ~0.000 & ~0.162 & 0.13 \\
$[$Mg/Fe$]$~II &  0.000 & ~0.052 & -0.056 & -0.004 & ~0.144 & 0.09 \\
\hline
\end{tabular}
\label{t:tab6}
\end{table} 

\begin{table*}[htb]
\centering
\caption[]{Abundances for BHB stars (only available in electronic form)}
\begin{tabular}{lccccccccccccc}
\hline
Star &EW(Y)&Y$_{NLTE}$&$[$N/Fe$]$&err&$[$O/Fe$]$&err&$[$Na/Fe$]$&err&$[$Mg/Fe$]$&err\\
\hline
36201 & 27.0 & 0.36 & 1.04 & 0.14 & 0.25 & 0.16 & 0.32 & 0.21 & 0.62 & 0.04 \\
36473 &      &      & 0.95 & 0.14 & 0.63 & 0.14 & 0.14 & 0.18 & 0.56 &      \\
37378 &      &      & 0.84 & 0.20 & 0.47 & 0.11 & 0.10 & 0.25 & 0.51 & 0.13 \\
44646 &      &      & 0.75 & 0.17 & 0.50 & 0.03 & 0.01 & 0.12 & 0.47 & 0.21 \\
45828 &      &      & 0.80 & 0.17 & 0.44 & 0.09 & 0.14 & 0.18 & 0.49 & 0.06 \\
47863 &      &      & 0.94 & 0.10 & 0.26 & 0.17 & 0.18 & 0.01 & 0.60 &      \\
48268 &      &      & 0.81 & 0.13 & 0.41 & 0.14 & 0.12 & 0.12 & 0.47 &      \\
49440 &      &      & 0.77 & 0.20 & 0.29 & 0.09 & 0.09 & 0.11 & 0.52 & 0.27 \\
54070 &      &      & 0.75 & 0.16 & 0.45 & 0.18 &-0.01 & 0.11 & 0.47 & 0.26 \\
54670 &      &      & 0.79 & 0.08 & 0.60 & 0.14 &-0.04 & 0.10 & 0.66 &      \\
55259 &      &      & 0.80 & 0.16 & 0.39 & 0.19 &-0.01 & 0.14 & 0.55 & 0.02 \\
56422 & 18.4 & 0.38 & 1.04 & 0.16 & 0.15 & 0.09 & 0.07 & 0.07 & 0.64 & 0.03 \\
61280 &      &      & 0.77 & 0.11 & 0.44 & 0.12 &-0.13 & 0.05 & 0.32 &      \\
63763 &      &      & 0.83 & 0.17 & 0.35 & 0.21 & 0.10 & 0.07 & 0.42 &      \\
68779 &      &      & 0.78 & 0.15 & 0.26 & 0.10 &-0.14 & 0.13 & 0.57 & 0.03 \\
71676 &      &      & 0.85 & 0.13 & 0.40 & 0.19 &-0.04 & 0.07 & 0.52 &      \\
77469 &      &      & 0.86 & 0.13 & 0.33 & 0.13 &-0.11 & 0.09 & 0.49 & 0.19 \\
\hline
\end{tabular}
\label{t:tab4}
\end{table*} 

\begin{table*}[htb]
\centering
\caption[]{Abundances for RHB stars (only available in electronic form)}
\begin{tabular}{lcccccccccccc}
\hline
Star &$[$Fe/H$]$&err&$[$O/Fe$]$&err&$[$Na/Fe$]$&err&$[$Mg/Fe$]$&$[$Si/Fe$]$&err&$[$Ca/Fe$]$&$[$Ni/Fe$]$&$[$Ba/Fe$]$\\
\hline
32170 & -1.11 & 0.24 & 0.45 & 0.12 & 0.03 & 0.11 & 0.59 & 0.45 & 0.09 & 0.78 & 0.18 & 0.27 \\ % 6.6
34102 & -1.22 & 0.22 & 0.53 & 0.10 & 0.03 & 0.11 & 0.54 & 0.51 & 0.04 & 0.69 &-0.09 & 0.58 \\ % 8.7
34668 & -1.26 & 0.24 & 0.53 & 0.05 & 0.14 & 0.15 & 0.52 & 0.63 & 0.09 & 0.81 & 0.26 & 0.59 \\ % 7.0
41649 & -1.23 & 0.22 & 0.46 & 0.02 & 0.14 & 0.07 & 0.55 & 0.64 & 0.09 & 0.85 &-0.13 & 0.81 \\ %10.0
42553 & -1.34 & 0.20 & 0.55 & 0.09 & 0.15 & 0.05 & 0.60 & 0.70 & 0.09 & 0.70 &      & 0.90 \\ % 8.2
42857 & -1.25 & 0.20 & 0.62 & 0.16 & 0.17 & 0.13 & 0.53 & 0.48 & 0.09 & 0.82 &      & 0.63 \\ %15.7
43379 & -1.21 & 0.21 & 0.60 & 0.03 & 0.09 & 0.15 & 0.44 & 0.64 & 0.13 & 0.50 &-0.06 & 0.83 \\ % 9.0
44426 & -1.08 & 0.19 & 0.80 & 0.14 & 0.20 & 0.12 & 0.50 & 0.58 & 0.11 & 0.69 &-0.27 & 0.67 \\ %12.0
46519 & -1.19 & 0.18 & 0.52 & 0.13 & 0.13 & 0.06 & 0.47 & 0.50 & 0.08 & 0.77 &-0.18 & 0.73 \\ % 8.0
48528 & -1.18 & 0.23 & 0.51 & 0.04 & 0.27 & 0.04 & 0.44 & 0.62 & 0.09 & 0.86 &-0.17 & 0.81 \\ % 9.3 x
49317 & -1.26 & 0.12 & 0.43 & 0.10 & 0.19 & 0.15 & 0.55 & 0.72 & 0.07 & 1.04 &-0.13 & 0.94 \\ % 9.2
51670 & -1.21 & 0.15 & 0.42 & 0.08 & 0.24 & 0.13 & 0.36 & 0.57 & 0.13 & 0.89 &-0.27 & 0.93 \\ %11.6
52012 & -1.23 & 0.23 & 0.65 & 0.08 & 0.08 & 0.26 &      & 0.60 & 0.12 & 0.99 &      & 0.71 \\ %12.3
54960 & -1.20 & 0.24 & 0.57 & 0.20 & 0.17 & 0.13 & 0.53 & 0.55 & 0.11 & 0.90 &-0.04 & 0.83 \\ % 8.9
56265 & -1.35 & 0.19 & 0.50 & 0.06 &-0.10 & 0.03 & 0.58 & 0.49 & 0.10 & 0.76 & 0.23 & 1.03 \\ % 6.2
58421 & -1.18 & 0.19 & 0.61 & 0.18 & 0.13 & 0.05 & 0.46 & 0.55 & 0.04 & 0.78 & 0.07 & 0.88 \\ %13.8
58511 & -1.28 & 0.22 & 0.51 & 0.14 & 0.01 & 0.17 & 0.49 & 0.64 & 0.08 & 0.70 &      & 0.61 \\ %13.4
58978 & -1.19 & 0.16 & 0.48 & 0.12 & 0.30 & 0.07 & 0.54 & 0.57 & 0.14 & 0.67 &-0.33 & 0.70 \\ %11.3
61405 & -1.17 & 0.18 & 0.49 & 0.14 & 0.09 & 0.10 & 0.52 & 0.62 & 0.06 & 0.90 & 0.18 & 0.80 \\ %10.0
61969 & -1.31 & 0.12 & 0.49 & 0.14 & 0.13 & 0.09 & 0.54 & 0.73 & 0.09 & 0.87 & 0.25 & 0.69 \\ % 7.2
63655 & -1.10 & 0.25 & 0.37 & 0.07 & 0.08 & 0.04 & 0.44 & 0.59 & 0.05 & 0.70 & 0.04 & 0.62 \\ % 6.2
65270 & -1.31 & 0.18 & 0.48 & 0.14 & 0.15 & 0.05 & 0.56 & 0.64 & 0.15 & 0.72 &      & 0.72 \\ % 8.7
65307 & -1.19 & 0.20 & 0.57 & 0.05 & 0.10 & 0.07 & 0.48 & 0.63 & 0.10 & 0.91 &-0.03 & 0.61 \\ % 7.5
65835 & -1.31 & 0.14 & 0.59 & 0.17 & 0.17 & 0.23 & 0.58 & 0.69 & 0.17 & 1.06 & 0.35 & 1.05 \\ % 5.9
66610 & -1.14 & 0.18 & 0.50 & 0.08 & 0.04 & 0.07 & 0.44 & 0.51 & 0.04 & 0.75 & 0.08 & 0.70 \\ %11.9
67352 & -1.28 & 0.17 & 0.62 & 0.11 & 0.03 & 0.15 & 0.48 & 0.62 & 0.21 & 0.70 & 0.14 & 1.19 \\ %13.2
71268 & -1.21 & 0.15 & 0.63 & 0.11 & 0.07 & 0.06 & 0.52 & 0.61 & 0.05 & 0.83 &      & 0.84 \\ % 9.4 x
72060 & -1.21 & 0.26 & 0.85 & 0.15 & 0.15 & 0.19 & 0.37 & 0.56 & 0.04 & 0.99 & 0.06 & 0.61 \\ % 6.8
75691 & -1.41 & 0.28 & 0.61 & 0.03 &-0.04 & 0.21 & 0.54 & 0.53 & 0.10 & 0.82 & 0.18 & 0.80 \\ % 5.2
75716 & -1.13 & 0.23 & 0.41 & 0.08 &-0.02 & 0.09 & 0.37 & 0.57 & 0.14 & 0.74 & 0.05 & 0.50 \\ % 9.8 x
\hline
\end{tabular}
\label{t:tab5}
\end{table*} 

\begin{center}
\begin{figure}
\includegraphics[width=8.8cm]{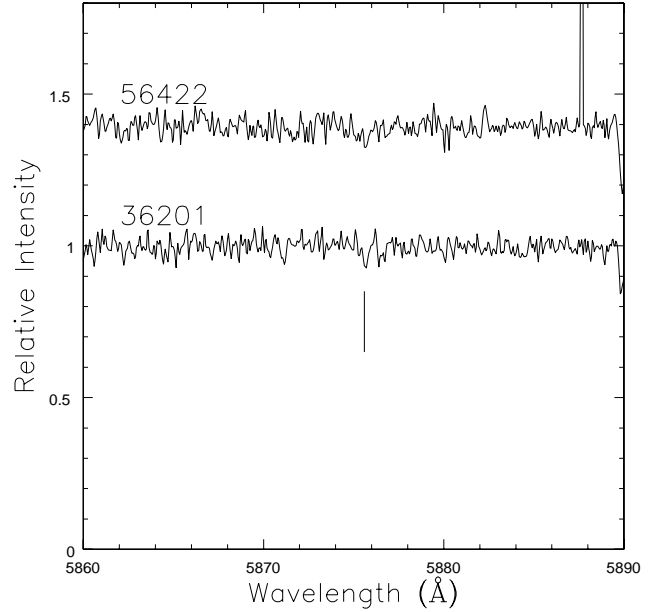}
\caption{Spectra in the region of the He~I line at 5876~\AA\ for stars \#36201
and \#56422. Spectra have been offset for clarity.}
\label{f:helines}
\end{figure}
\end{center}

%\begin{table*}[htb]
%\centering
%\caption[]{Average He abundances from the 5876~\AA\ line for selected clusters}
%\setlength{\tabcolsep}{1.5mm}
%\begin{tabular}{lcccccccccc}
%\hline
%Cluster &$[$Fe/H$]$&Rel. Age (1) &
%$\log{T_{\rm eff}}$(HB)&$\log{T_{\rm eff}}$(HB)&$\log{T_{\rm eff}}$(HB)&
%Ref. & N$_{\rm stars}$ & $\log{T_{\rm eff}}$&  $<Y>$& r.m.s. \\
%&&& Min & Median & Max &&& Range \\
%\hline
%NGC1851 & -1.18 &0.81 & 3.73 & 3.74 & 4.08 & 2 & 19 & 3.95$\div$4.06 &$0.297\pm 0.020$ & 0.088  \\ % Bluest stars
%NGC2808 & -1.18 &0.83 & 3.75 & 3.92 & 4.57 & 3 & 17 & 3.96$\div$4.06 &$0.336\pm 0.013$ & 0.052  \\ % Interm. pop.
%M~5     & -1.33 &0.85 & 3.76 & 3.89 & 4.18 & 4 & 15 & 3.95$\div$4.02 &$0.312\pm 0.017$ & 0.064  \\ % Interm. pop. 
%M~4     & -1.18 &0.97 & 3.72 & 3.76 & 4.04 & 5 &  6 & 3.95$\div$3.98 &$0.295\pm 0.011$ & 0.028  \\ % Bluest stars
%M~22    & -1.70 &1.06 & 3.82 & 3.97 & 4.22 & 6 & 29 & 3.95$\div$4.03 &$0.338\pm 0.014$ & 0.074  \\ % Metal-rich pop.
%NGC6752 & -1.55 &1.02 & 3.82 & 4.02 & 4.47 & 7 &  4 & 3.93$\div$3.94 &$0.252\pm 0.016$ & 0.031  \\ % Reddest stars
%\hline
%\end{tabular}
%\\
%1. From Gratton et al. (2010); 
%2. Gratton et al. (2012a); 
%3. Marino et al. (2013b);
%4. Gratton et al. (2013);
%5. Villanova et al. (2012);
%. This paper;
%7. Villanova et al. (2009)
%\label{t:tabhe}
%\end{table*} 

\subsection{Line list and notes on individual elements}

The abundance analysis done in this paper generally follows the scheme outlined in
previous papers of this series. In particular, most of the abundances we derived
assumed local thermodynamic equilibrium (LTE); however, non-LTE corrections were included for N (Przybilla \&
Butler 2001; see Gratton et al. 2012b), O (Takeda et al. 1997),
and Na (Mashonkina et al. 2000). The He abundances follow the prescriptions in Gratton et al.
(2014) and should be considered homogeneous to the non-LTE analysis by Marino et al. (2013).

The only peculiar point concerns the use of only two of the lines in the O~I triplet
at 7771-75~\AA. In fact, we found that in our spectra, the weakest line of the 
triplet at 7775.4~\AA\ falls very close to a quite strong telluric emission line
because of the combination of the cluster velocity and Earth's motion around the barycentre
of the solar system (the two observations with HR19 were obtained at short cadence). 
To avoid uncertain corrections, we preferred to remove this line from our analysis.

\subsection{Sensitivity of abundances on the atmospheric parameters}

The sensitivity of abundances to the adopted values for the atmospheric parameters is given 
in Table~\ref{t:tab6}. It was obtained as usual by changing each parameter separately and
repeating the abundance analysis. We also considered the contribution to the error
due to uncertainties in the equivalent widths, divided by the square root of the typical
number of lines used in the analysis. The values were computed for typical uncertainties 
in each parameter, as determined in Sect. 3.1. Results are given for an RHB and a BHB
star. 

\begin{table}[htb]
\centering
\caption[]{Average abundances for BHB and RHB stars}
\setlength{\tabcolsep}{1.5mm}
\begin{tabular}{lcccc}
\hline
Parameter         & BHB  &r.m.s.&  RHB  &r.m.s.\\
\hline
$Y$               & 0.37 & 0.01 &  ..   & ..   \\
\hline
$[$Fe/H$]$        &   .. &  ..  & -1.22 & 0.07 \\ 
$[$N/Fe$]$~I      & 0.85 & 0.09 &  ..   & ..   \\
$[$O/Fe$]$~I      & 0.39 & 0.13 & ~0.53 & 0.09 \\
$[$Na/Fe$]$~I     & 0.05 & 0.12 & ~0.13 & 0.09 \\
$[$Mg/Fe$]$~I     &  ..  &  ..  & ~0.51 & 0.06 \\
$[$Mg/Fe$]$~II    & 0.52 & 0.08 &  ..   & ..   \\
$[$Si/Fe$]$~I     &  ..  &  ..  & ~0.60 & 0.08 \\
$[$Ca/Fe$]$~I     &  ..  &  ..  & ~0.81 & 0.13 \\
$[$Ni/Fe$]$~I     &  ..  &  ..  & -0.01 & 0.20 \\
$[$Ba/Fe$]$~II    &  ..  &  ..  & ~0.75 & 0.17 \\
\hline
\end{tabular}
\label{t:tab7}
\end{table} 

\begin{center}
\begin{figure}
\includegraphics[width=8.8cm]{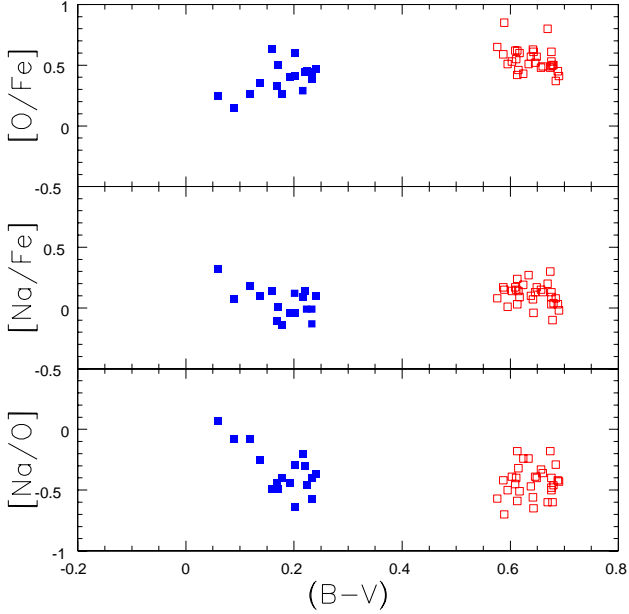}
\caption{Run of the [O/Fe] (upper panel), [Na/Fe] (middle panel), and of the [Na/O]
abundance ratios (lower panel) as a function of the $B-V$ colour.}
\label{f:ona}
\end{figure}
\end{center}

\begin{center}
\begin{figure}
\includegraphics[width=8.8cm]{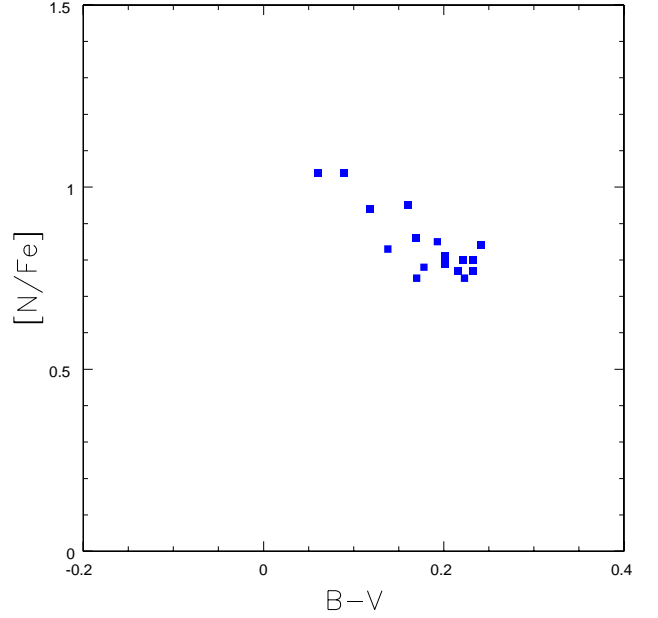}
\caption{Run of the N  as a function of the $B-V$ colour.}
\label{f:nitrogen}
\end{figure}
\end{center}

\begin{center}
\begin{figure}
\includegraphics[width=8.8cm]{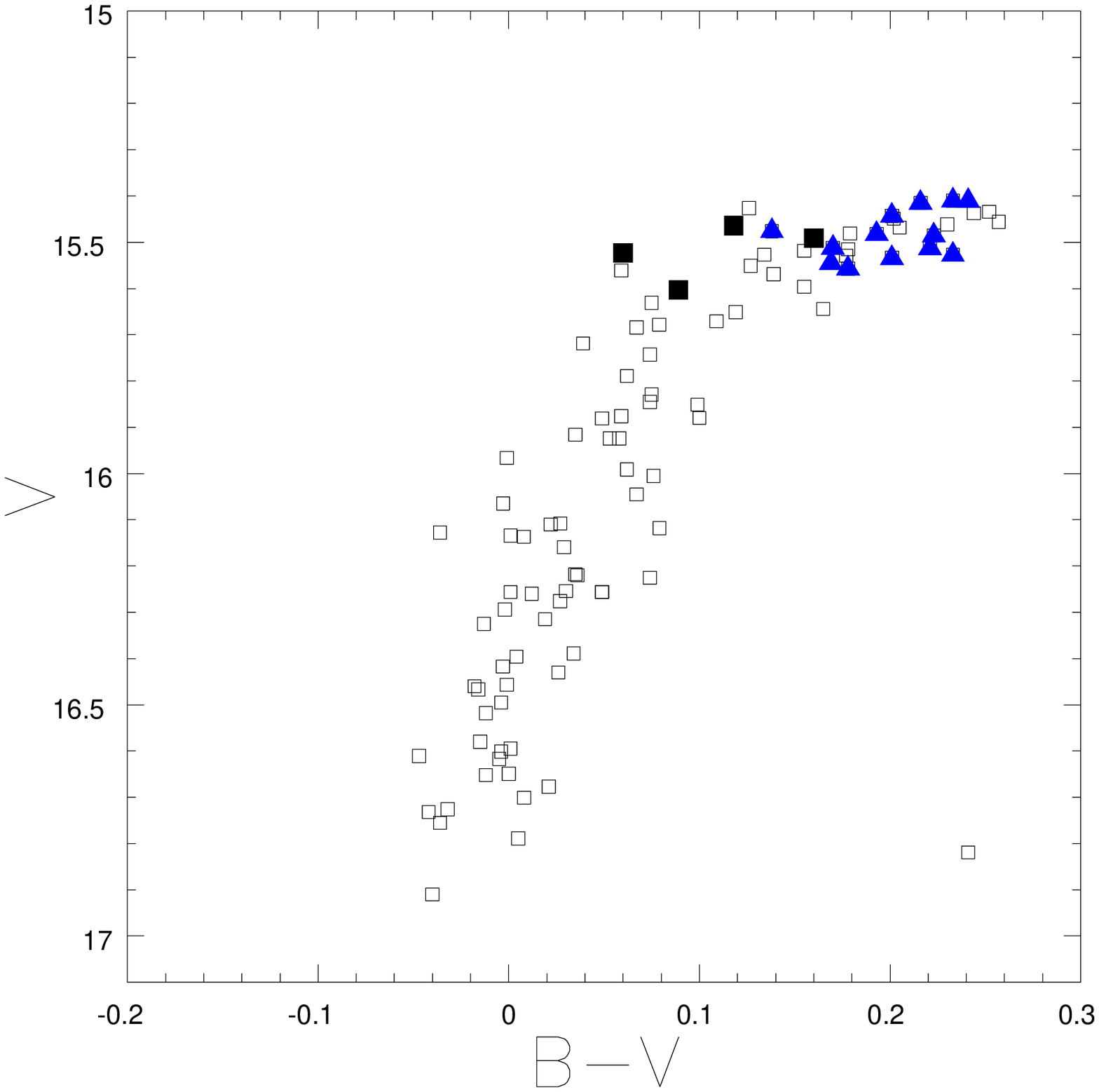}
\caption{Zoom on the BHB of NGC~6723; filled symbols are used for O-rich/Na-poor/N-poor (blue
triangles) and O-poor/Na-rich/N-rich stars (black squares). Open squares are for stars not
observed in this paper. }
\label{f:hb}
\end{figure}
\end{center}

\section{Results}

\subsection{Metal abundance}

Average abundances for BHB and RHB stars are given in Table~\ref{t:tab7}, along with
the r.m.s. scatter around these mean values. The star-to-star scatter generally agrees
fairly well with expectations based on uncertainties in the atmospheric parameters
and in the $EW$s, as given in the last column of Table~\ref{t:tab6}. This suggests
that most stars share the same chemical composition. The average Fe abundance (only
derived for RHB stars) is [Fe/H]=$-1.22\pm 0.01$\, with an r.m.s. dispersion for
individual stars of 0.07~dex. This error bar only includes the statistical uncertainties;
systematic effects are probably stronger (about 0.08~dex) and are mainly dominated by errors 
in the \teff . This value of the Fe abundance agrees well with the literature determinations
listed in the Introduction, in particular with the only other high-dispersion study by
Fullton \& Carney (1996). The $\alpha-$elements Mg, Si, and Ca are all overabundant,
again in agreement with the earlier study by Fullton \& Carney (1996).

\subsection{p-capture elements}

For N, O, and Na, all elements whose abundances are affected by p-capture reactions
and that are the focus of our analysis, our results indicate that most stars share the
same high O and low Na abundances, typical of first generation stars in GCs (Gratton et al. 2012a).
This result was expected for the RHB stars; however, almost all BHB stars also share
this abundance pattern (see Fig.~\ref{f:ona}). Only four among the warmest BHB stars
are possibly depleted in O and enriched in Na compared to the remaining stars.
This evidence is much strengthened by the consideration that these four stars are
also enriched in N, with values of [N/Fe]$\sim 1$~dex compared to the lower value of
[N/Fe]$\sim 0.8$~dex observed in the other 13 BHB stars (see Fig.~\ref{f:nitrogen}). It is interesting to note that
these stars are on the bluest and brightest side of a small gap that is possibly visible
on the HB of NGC6723 at \teff$\sim 9000$~K (see Fig.~\ref{f:hb}). While this gap does not
appear to be very robust from a statistical point of view (see analysis of possible gaps 
found in other cases, e.g. Catelan et al. 1998), we used it throughout this paper as a 
convenient reference point to separate HB stars according to their chemical composition.
Whether a real gap is actually present at this this colour or not, we found that there is
segregation of first and second generation stars on the HB of NGC~6723 at this colour.
This segregation agrees with the hypothesis that the
chemical abundances are important factors in determining the distribution of stars along the HB.
On the other hand, the wide span in colours ($0.07<(B-V)<0.63$) for stars sharing a similar 
chemical composition suggests that some spread in mass can exist even at constant chemical
composition.

\subsection{Helium}

We tried to measure He abundances for the two warmest stars we observed in NGC~6723
(\#36201 and \#56422); the remaining stars are too cool to show evidence of helium lines. 
The method used is described in detail in
Gratton et al. (2014). The error bar to be attached is the sum of errors due to
atmospheric parameters and equivalent widths; they were obtained by repeating the
analysis with values at the edge of the error bars for these quantities.
We obtained very high values of the helium abundance
($Y\sim 0.37$), with similar values derived for both stars. However, the error 
bar to be attached to these He abundance determinations is large (about $\pm 0.08$), given 
the uncertainties in the $EW$s and in the \teff\ (these last
are larger than usual for these warm stars). The large error bar allows values of the
He abundance only slightly higher than the value expected from Big Bang nucleosynthesis 
($Y\sim 0.25$: Cyburt 2004) and a modest enrichment due to first dredge-up 
($\Delta Y\sim 0.015$: Sweigart 1987). We therefore recommend caution when using these He 
abundances.

\begin{table}[htb]
\centering
\caption[]{Properties of RHB stars}
\setlength{\tabcolsep}{1.5mm}
\begin{tabular}{lccccc}
\hline
NGC   &$[$Fe/H$]$ &$[$(Mg,Si)/Fe$]$&$[$M/H$]$&Relative&$\theta_{\rm eff}$(RHB)\\
      &           &                &         &Age     &                       \\
\hline
~104 & -0.76 & 0.26	& -0.70 & 0.99 & 1.008 \\
1851 & -1.18 & 0.34	& -1.10	& 0.79 & 0.940 \\ 
2808 & -1.18 & 0.30 & -1.10	& 0.88 & 0.925 \\
5904 & -1.27 & 0.24	& -1.21	& 0.86 & 0.876 \\
6723 & -1.22 & 0.55	& -1.08	& 1.02 & 0.904 \\ 
\hline
\end{tabular}
\label{t:tabrhb}
\end{table} 

\section{Discussion and conclusions}

\subsection{Comparison with other clusters}

We may compare the abundances we obtained for NGC~6723 with those we obtained for
other GCs in this series of papers; since the methods are very similar, we expect that the analysis
can be considered homogeneous. We may perform a test on the RHB stars, which constitute
a more homogeneous sample of stars. In fact, because we expect that when the analysis is limited to
stars belonging to the FG, they should have an original He abundance close to that of the Big 
Bang (Cyburt 2004) and their average effective temperatures should be determined 
by a combination of metal abundance, age, and mass loss. We observed five clusters with RHB stars:
47 Tuc ([Fe/H]=-0.76; Gratton et al. 2013),  M5 ([Fe/H]=$-1.27$; Gratton et al. 
2013), NGC~1851 ([Fe/H]=$-1.18$; Gratton et al. 2012b), and NGC~2808 ([Fe/H]=$-1.18$; Gratton 
et al. 2011), in addition to NGC~6723 ([Fe/H]=$-1.22$; this paper). We note that three
of these clusters (M5, NGC~1851 and NGC~2808) are younger than the two others according 
to Mar\'in-Franch et al. (2009). We also note that
all RHB stars in M5, NGC~1851, and NGC~6723 have compositions compatible with their being
FG stars; however, only the reddest stars in 47~Tuc and NGC~2808 are bona-fide FG stars,
the bluest ones are enriched in Na and depleted in O.
Table~\ref{t:tabrhb} collects the most important properties of the RHBs of these clusters.
Abundances were from this series of papers, where [(Mg,Si)/Fe] is the average
of [Mg/Fe] and [Si/Fe] values. Note that we obtained an overall metal abundance by 
combining the abundances of Fe, Mg, and Si with the rule [M/H]=[Fe/H]+0.25~[(Mg,Si)/Fe]
\footnote{This formula is different from that of Straniero \& Chieffi (1991); its
justification here is simply that it gives a better fit to the data.}.

Relative ages are the average of the values obtained by Mar\'in-Franch et al. (2009)
using the metallicities reported by Zinn and West (1984) and Carretta and Gratton (1997). Finally,
$\theta_{\rm eff}$(RHB)=5040/\teff(RHB), where \teff(RHB) is the average temperature
for those RHB stars that have the typical composition of FG stars. Although data are available
for only five clusters, a bivariate analysis shows
a very good correlation between $\theta_{\rm eff}$(RHB) and a combination 
of [M/H] and age (Pearson linear correlation coefficient of r=0.98).
This result does not depend critically on the particular set of ages we adopted
for globular clusters; in fact, we derive the same Pearson linear correlation coefficient 
of r=0.98 when using ages from VandenBerg et al. (2013), for example.

This comparison not only supports our analysis, but also suggests that the
FG stars we observed on the BHB of M5 and NGC~6723 possibly occupy this location
on the HB because they lost more mass than normal while evolving along the red giant
branch. This is a possible clue for interpretating of the basic mechanism
involved. 

The element-to-element abundances agree very well 
among different clusters for O and Na. NGC~6723 seems to have an excess
of $\alpha$-elements larger than that observed in the other clusters. While this
might be related to its greater age, there is no one-to-one correlation between
age and excess of $\alpha$-elements because NGC~1851, the youngest cluster in our
sample, is the second-richest in $\alpha$-elements. The Ca abundances for NGC~6723
are very high. Part of this trend seems to be a systematic effect in our analysis of RHB
stars, which systematically produces values of [Ca/Fe]$\sim 0.2$~dex higher than
those obtained from studies of RGB stars (Carretta et al. 2010), and for 
[Mg/Fe] and [Si/Fe] in the same RHB stars, with a trend for larger differences in
warmer stars. Since our Ca abundances are based on rather strong lines with very 
accurate line data, this difference might be due to non-LTE effects
\footnote{Statistical equilibrium computations for neutral calcium lines have been produced
by Mashonkina et al. (2007). These authors give non-LTE abundance corrections for one of the lines used in the 
current analysis (5857~\AA), but not for the other one (6122~\AA), but they give
results for the strongest line of the same multiplet at 6162~\AA. Unfortunately, the
range of surface gravity they considered does not extend to values appropriate for
the RHB, so that some extrapolation would be needed. However, the non-LTE corrections 
they found are small, and the extrapolations produce values $<0.1$~dex.}
However, even if corrected for this effect, the
[Ca/Fe] ratio found for NGC~6723 stars is still high; this might explain
the high metal abundance inferred from applying the $\Delta S$\ method to the
RR Lyrae ([Fe/H]$\approx -0.7$: Smith 1981) and the quite high value listed
by Zinn \& West (1984). A similar suggestion was made by Smith \& Hesser (1986).

\begin{center}
\begin{figure}
\includegraphics[width=8.8cm]{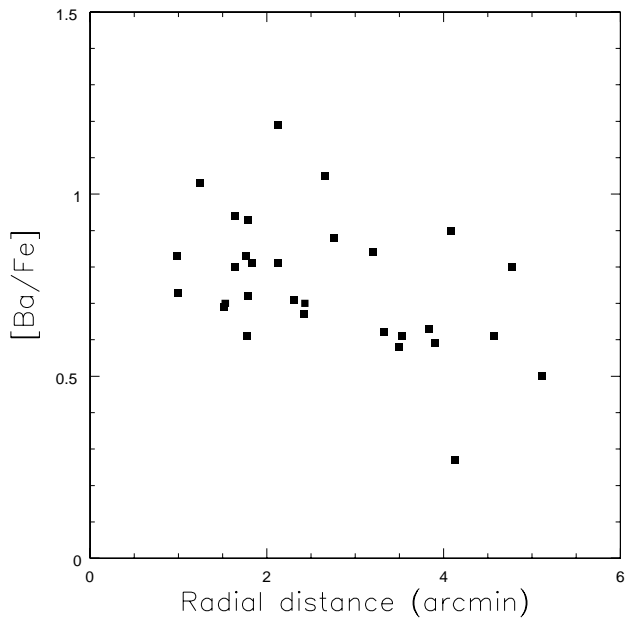}
\caption{[Ba/Fe] values as a function of projected distance from center of the cluster.}
\label{f:figba}
\end{figure}
\end{center}

\subsection{Ba abundances}

We obtained a very high abundance of Ba in NGC6723 ([Ba/Fe]=0.75). Part of this
high value can be systematic because we also obtained quite high abundances of Ba
for the other clusters we observed. However, when compared with results for RGB 
stars for the other clusters in this series (Lucatello et al., in preparation), the systematic 
difference is small, if there is any at all ($0.02\pm 0.16$). We notice that high abundances of Ba 
have also been obtained for other moderately metal-rich old GCs: [Ba/Fe]=0.63, 
0.46 and 0.45 for NGC~6121 (=M4: [Fe/H]=$-1.17$), NGC~6171 (=M107: [Fe/H]=$-1.03$), and 
NGC~6752 ([Fe/H]=$-1.56$) (barium abundances are from D'Orazi et al. 2010, 
and Lucatello et al., in preparation, and [Fe/H] values from Carretta et al. 2009b).

However, the most intriguing result we obtained concerning Ba is the 
correlation between the [Ba/Fe] values and the projected distance from cluster
center (see Fig.~\ref{f:figba}). The Pearson linear correlation coefficient is 0.46
over 30 stars, which has a probability lower than 0.5\% to be a random effect. The
abundance gradient is not very strong, and RHB stars close to the center of the
cluster have Ba abundances that are $\sim 0.2$~dex higher than those at the cluster
periphery. This effect is not due to any individual star, and Ba abundances do not 
correlate with the scatter of radial velocities. There is then no evidence that it
is due to mass transfer episodes in close binaries (classical Ba-stars: McClure 1983;
McClure \& Woodsworth 1990). We
found no evidence for radial gradients in the abundances of any other elements,
nor any correlation of the Ba abundances with the atmospheric parameters. The
Ba abundance gradient seems, then, a real feature of NGC~6723.

Unfortunately, Ba is the only n-capture element we observed. Hence we cannot tell whether
this gradient is a consequence of s- or r-processes, which would lead to very
different interpretations. Were it due to the r-process, the case could be analogous
to that observed in M15 (Sneden et al. 1997, 2000) and M92 (Sobeck et al. 2011).
The astrophysical site for the r-process is still unknown (see Arnould et al. 2007
for a discussion) with either some special kind of core-collapse supernovae (SNe)
(possibly related to gamma-ray bursts: Fujimoto et al. 2008) or merging neutron stars being 
proposed. The event is probably quite rare to explain the large variance in the 
abundances of the r-process elements found in extremely metal-poor stars (Gilroy et al. 
1988; McWilliam et al. 1995a, 1995b; Roederer et al. 2010) but not too rare to 
avoid too much a scatter (Argast et al. 2004). Regardless its nature, the event probably 
occurred within the proto-cluster when it was still very gas-rich and most likely 
occurred in the central regions, which may explain the radial abundance gradient in 
Ba observed in NGC~6723. On the other hand, because of the much longer relaxation 
time, it is more likely that a radial abundance gradient is preserved in NGC~6723 than 
in M15. The lowest Ba abundance of NGC~6723 is much higher than that of M15, requiring 
a more effective Ba production to produce sizable differences in the Ba abundance; but 
NGC~6723 is a smaller cluster and the star-to-star variations in Ba abundances are
smaller, which at least partly compensates for this difference. Since the total Ba content 
of NGC~6723 is $\sim 10^{-5}~M_\odot$, this is roughly the amount of Ba that must have been
produced by this r-process event (if it was a single episode); note that this is quite a
high value for 
core-collapse SNe (Argast et al. 2004), while it is easily achieved in neutron star mergers (see Argast et al. 
2004) and collapsars (Fujimoto et al. 2008). On the other hand, no sizeable abundance
difference has been found for elements other than Ba: this limits the total number of
SNe that could have exploded in this phase. Since there are $\sim 10~M_\odot$ of O
within NGC~6723, the total number of polluting SNe during this phase was probably 
small, no more than a few (roughly the same numbers also hold for M~15). We
note that since in the Sun the Ba/O mass ratio is about $10^{-6}$, and about 15\%
of the solar Ba is due to the r-process (Burris et al. 2000), the Ba/O overproduction 
does not need necessarily to be extreme. Hence this scenario is well feasible. We 
finally mention that it is unclear whether applying this scenario to M~15, M~92,
and NGC~6723 requires any special connection between the r-process event and GCs. A 
dedicated search for the spread in the r-process element abundances in GCs might help to 
clarify this question.

On the other hand, the Ba production required to explain the observed gradient might
also be due to the s-process. The main candidate in this case is the main component
because we do not have evidence for variations in lighter elements expected from
production by the weak component in massive stars. In this case, multiple populations
should indeed be present even among the RHB stars of NGC~6723, with age differences
so large that pollution is expected to be due to stars too small to have a significant hot
bottom burning. The age spread probably is of hundred million years, possibly similar
to that proposed to explain the extended turn-off of several intermediate-age clusters
in the Magellanic Clouds (Mackey \& Broby Nielsen 2007; Milone et al. 2009). However, in this case it appears 
difficult to avoid variations in the total CNO content and split SGB, such as those 
observed in NGC~1851 (Milone et al. 2008; Cassisi et al. 2008) and M22 (Marino et al.
2009). There is no evidence for anything similar in NGC~6723, and we note
that there is no radial gradient in N abundances, for instance. However, these 
arguments are circumstantial and not strong enough to completely discard this hypothesis.
Only determination of the abundances for elements with high r-fraction 
abundances (e.g., Eu) may clarify this question.

We finally add a caveat about the previous discussion: the original mass
of globular clusters was certainly higher than the current one, by an amount that is
probably different from cluster to cluster and can even be large (see e.g. D'Antona \& Caloi 
2004, and Lamers et al. 2010, MNRAS, 409, 305). All numbers given in
this section should therefore be considered with caution and at most as order-of-magnitude
estimates. The only conclusion is that at present there is no obvious reason for
either an r- or an s-process explanations of the observed scatter in Ba abundances in
NGC~6723.  

\subsection{Distribution of stars along the HB}

When discussing our results, we first note that our sample is not exactly representative
of the whole HB population of NGC~6723. We may use the photometry by Lee et al. (2014), which
extends over the whole cluster, to derive the total number of stars in different portions of
the HB. Field contamination is a problem for RHB stars, while we may safely neglect it for the
BHB ones. To reduce this concern, we will only used counts of stars 
within 7.5 arcmin; this limit approximately corresponds to the projected
distance from the center of the farthest radial velocity member star in our sample. We 
have observed 30 RHB stars over a total of about 132; this last
total was obtained assuming that the percentage of contaminants is as high in the whole photometric 
sample as it is in our spectroscopic one within 7.5 arcmin (30 out of 34 our candidate RHB stars 
within this distance are cluster members). We also observed 4 RR Lyrae stars out of 42;
15 out of 35 BHB stars cooler than the gap that we tentatively found at $\sim 9000$~K ($B-V\sim 0.1$), 
and that may roughly separate the first and second generation along the HB of NGC~6723; and two out of 
68 stars warmer than this gap. We note, however, that we also found a N- and Na-excess and an O
deficiency for two stars slightly cooler than the possible gap; these stars are
slightly brighter than most of the stars with the same colour. Their location
on the HB suggests that they are evolved objects that started their HB evolution on the
bluer side of the tentative gap. We then combined these stars with the group of stars bluer than the
gap. For simplicity we call this last group extreme-BHB below to separate 
them from the redder stars on the intermediate-BHB that are cooler than this gap. 

According to our determinations of the O, Na, and N abundances, RHB and intermediate-BHB stars
of NGC~6723 all belong to the first generation because they are O-rich, Na-poor and N-poor. 
While we have not analysed any of the RR Lyrae stars, we expect that they share the composition
of the red and intermediate-BHB stars that bracket them in colours. Hence a total of 132+42+33=207
HB stars belong to the first generation. On the other hand, the only O-poor, Na-rich, and N-rich
stars we found in NGC~6723 are the four extreme-BHB stars (these four stars are among the coolest
extreme-BHB stars). These are the only stars we observed in NGC~6723 that probably belong to the 
second generation. Even if we assume that all the extreme-BHB stars belong to the second generation
(a reasonable assumption), the total of second-generation stars on the HB of
NGC6723 is 70 stars at most. Although lifetimes of the extreme-BHB stars might be slightly
different from those of the remaining HB stars and there is some statistical uncertainty, related
in particular to de-contamination of RHB stars from field interlopers, the conclusion that 
second-generation stars make up at most only a small fraction (70 out of 277, that is, about 
one fourth) of the total stars of NGC~6723 seems to us straightforward. This low fraction of 
second-generation stars distinguishes this cluster from many others observed (see, e.g., the 
compilation by Carretta et al. 2009a) where the second generation usually dominates the 
first generation and typically makes up some two thirds of
the cluster population: normalized to the first generation, the second generation of NGC~6723
is about $<1/8$\ that of a "normal" GC. The large variation of the first-to-second generation ratios
indicated by these numbers should provide a caution about, e.g., the use of clusters in the Fornax
dwarf galaxy to confirm or dismiss scenarios for internal pollution in GCs (Larsen et al. 2012;
Bastian et al. 2013); on the other
hand, it should also caution us concerning arguments about the fraction of halo stars that
formed in GCs from the population of luminous clusters surveyed so far to the total initial
population of globular clusters (Gratton et al. 2012b). We furthermore note that the low
central concentration of NGC~6723 may be somehow connected to this peculiarity, although
other clusters with a concentration even lower than NGC~6723, such as NGC~288 and NGC~6809, have
more normal fractions of FG and SG stars. We finally note that NGC~6723 is not included in the
list of clusters suspected to predominantly host FG stars by Caloi \& D'Antona (2011).

\begin{center}
\begin{figure}
\includegraphics[width=8.8cm]{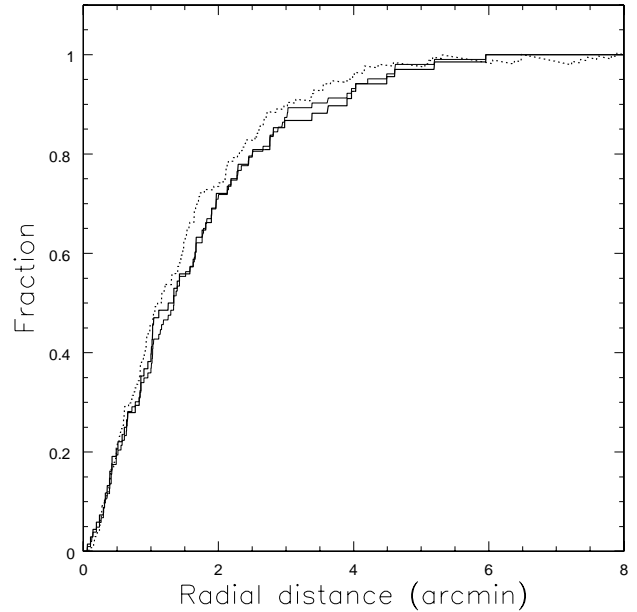}
\caption{Cumulative radial distribution for stars on different parts of the HB
of NGC~6723. Thin solid line is for stars on the whole BHB, the thick solid line
is for stars on the extreme BHB, and the dotted line is for stars on the RHB.}
\label{f:cumdist}
\end{figure}
\end{center}

There are several indications that in many clusters second-generation stars are much more centrally
concentrated than first-generation stars (see Gratton et al. 2012a). To test this point for NGC~6723, 
we plot in Fig.~\ref{f:cumdist}
the cumulative radial distribution of extreme-BHB stars and compare them with those of other
stars on the HB of this cluster. There is no evidence for any difference in the two distributions.
On the other hand, we note that NGC~6723 is a cluster with low central concentration, which
might be related to the unusual predominance of first-generation stars because FG stars
are typically less concentrated than SG stars.

\begin{center}
\begin{figure}
\includegraphics[width=8.8cm]{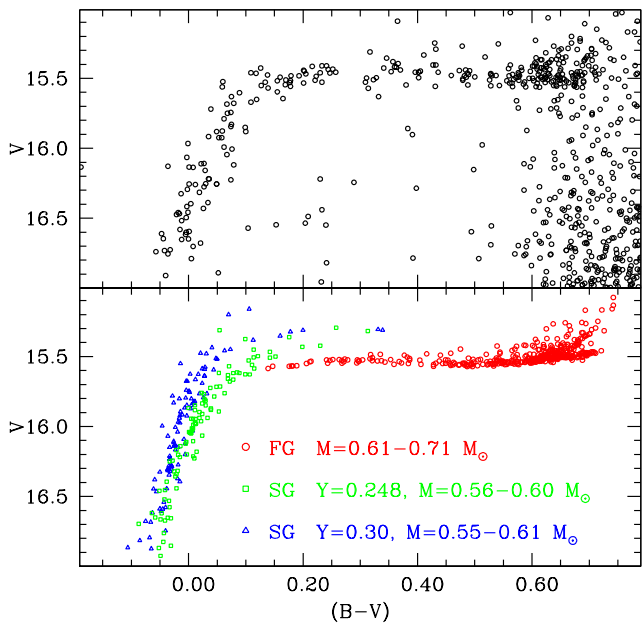}
\caption{Comparison between a synthetic (lower panel) and an observed
(upper panel) colour-magnitude diagram for the horizontal branch of
NGC~6723. In the lower panel, red circles are FG (Y=0.248) stars
(mass range 0.61-0.71~$M_\odot$),
while green triangles and blue open squares are SG with different values of the He
abundance: Y=0.248 and Y=0.30. The mass range for the
SG stars is 0.55-0.61~$M_\odot$\ in the first case and 0.56-0.60~$M_\odot$\
in the second. }
\label{f:comp1}
\end{figure}
\end{center}

\begin{center}
\begin{figure}
\includegraphics[width=8.8cm]{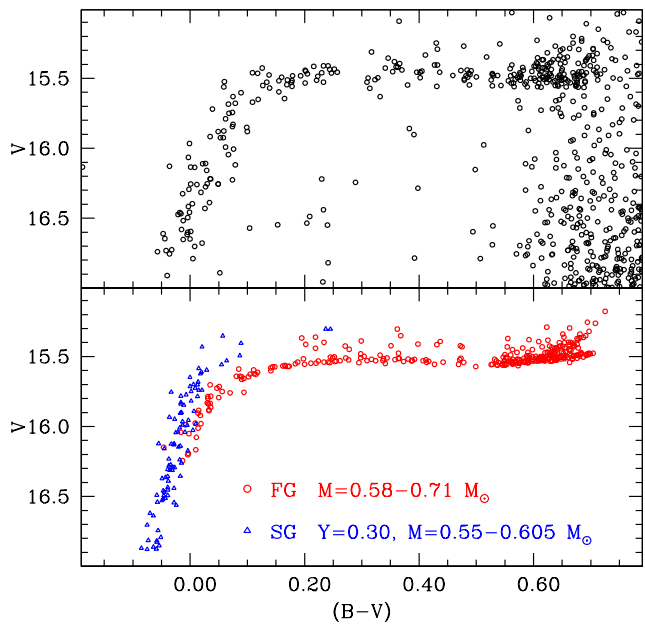}
\caption{Same as Fig.~\ref{f:comp1}, but in this case different ranges in
mass were adopted: 0.58-0.71~$M_\odot$\ for the FG (red open circles) and 0.55-0.605~$M_\odot$\
for the SG (Y=0.30) blue open triangles. }
\label{f:comp2}
\end{figure}
\end{center}

The other interesting fact concerning the HB of NGC~6723 is the wide distribution in colours of
first-generation stars. To quantify this spread in colour as a spread in mass and to discuss the
relation between first and second generation in NGC~6723, we compared the observed colour-magnitude
diagram with synthetic HBs obtained using the same tools as in Gratton et al. (2012b, 2013, 2014) 
and described in more detail in Salaris et al. (2008) and Dalessandro et al. (2011). We used the HB 
evolutionary tracks with [Fe/H]=$-1.25$\ from the BaSTI database (Pietrinferni et al. 2006),
interpolating among the $\alpha-$enhanced BaSTI models to determine HB tracks
for various values of the helium content Y. We adopted a distance modulus $(m-M)_V=14.84$\ and 
$E(B-V)=0.05$\ from Harris (1996, 2010 edition). For simplicity, we adopted discrete
values for the chemical composition; this assumption is not supported by strong
evidence for this particular cluster, but it may be a good approximation for others. 
However, this assumption is not essential in the following discussion.

We performed several simulations; in all cases we assumed that the helium-rich SG 
contains one fourth of the stars, as previously discussed. However, this exact value is 
not crucial here. Figure~\ref{f:comp1} contains a first comparison. In this case, 
we assumed that the FG stars (red circles) have a standard Big Bang He abundance (Y=0.248) and a
uniform distribution in mass over the range 0.61-0.71~$M_\odot$. Initially,
we assumed that SG stars are He-rich (Y=0.30) and have a uniform distribution in
mass over the range 0.55-0.61~$M_\odot$ (blue circles in Fig.~\ref{f:comp1}); this
value for Y is close to the lower extreme of the observational range. 
While the overall range in colours is well covered by this simulation, we note that 
with such assumptions there is quite a large discontinuity in luminosities ($\sim 0.2$~mag) 
at the transition between the two populations (at $B-V$=0.15). There is no such obvious 
discontinuity in the observed data. The discontinuity can be eliminated by assuming a
lower He abundance for the SG stars: to show this, we plot in this same figure
a synthetic SG computed with the range mass 0.56-0.60~$M_\odot$, but the same value
of Y as the FG one (Y=0.248: green points). However, this value of Y disagrees with 
the spectroscopic result.
% and with the expectation given by the high [Na/O] ratio observed in these stars.

A possible alternative is to assume that SG stars are He-rich, as indicated by spectroscopy,
but FG stars have a wider range in masses (0.58-0.71~$M_\odot$). A simulation performed with
this assumption is shown in Fig.~\ref{f:comp2}. We also slightly modified the range in mass
for the SG to 0.55-0.605~$M_\odot$, but this modification plays a secondary role here. While
SG are more luminous than FG stars at the same colour, there is now quite a substantial
overlap in colour between the two populations, so that there is no obvious discontinuity
at a colour of $B-V$=0.15. We note that in this simulation, the ratio between the number of
RR Lyrae variables and RHB stars is 0.27, close to the observed ratio (42/132), and that
the number of FG BHB stars is similar to that of RR Lyraes. This means that a uniform distribution
in masses for the FG stars over this wide range is an acceptableapproximation. If this 
interpretation of the HB of NGC~6723 is correct, the ratio between SG and FG stars is even 
lower than the value of one fourth we concluded in the previous section because some of the extreme 
BHB stars are FG stars.

%forse mi sono espresso male e devo correggere il testo, ma dal punto di
%vista dei rapporti di popolazione, che alcune (poche) stelle di extreme-BHB
%siano di FG invece che SG, non cambia molto. Non c'e' molta differenza se
%in questo amamsso la SG e' 1/4 o 1/5 del totale. Richiede chiaramente un ulteriore
%estensione del range di massa della FG, ma pure in questo caso la
%modifica e' da 0.10 a 0.13 Mo, che non e' un cambiamento sostanziale.
%Penso che siano dettagli rispetto ai risultati sostanziali ottenuti.

We conclude that to reproduce spectroscopic and photometric evidence, the range in
mass of FG - that is, of chemically homogeneous - stars in NGC~6723 is probably quite wide:
at least 0.10~$M_\odot$, and more likely about 0.13~$M_\odot$. This requires a wide distribution
in the mass lost by the stars along the RHB. We recall that a smaller but not negligible spread 
in mass loss among FG stars was also required to explain the distribution of stars along
the HB of M~5 (Gratton et al. 2013). This indicates that at least a fourth parameter (in
addition to metallicity, age, and helium abundance) is required to explain the HB of globular
clusters, in agreement with earlier suggestions based on other properties of HB stars (see, e.g.,
the discussion in Catelan 2009).

\begin{acknowledgements}
This publication makes use of data products from the Two Micron All Sky Survey, 
which is a joint project of the University of Massachusetts and the Infrared 
Processing and Analysis Center/California Institute of Technology, funded by the 
National Aeronautics and Space Administration and the National Science
Foundation. This research has made use of the NASA's Astrophysical Data System. 
This research has been funded by PRIN INAF ``Formation and Early Evolution of 
Massive Star Clusters". SL, AS, EC, SC, and AB acknowledge financial support from PRIN MIUR 2010-2011, 
project ``The Chemical and Dynamical Evolution of the Milky Way and Local Group 
Galaxies'' (PI F. Matteucci), prot. 2010LY5N2T. VD is an ARC Super Science Fellow.
We thank Jae-Woo Lee for sending us the full $BV$ photometric data they obtained for NGC~6723. 

\end{acknowledgements}

\end{document}